\documentclass[letterpaper,times]{IONconf}
\usepackage{url}

\usepackage{natbib}

\usepackage[hidelinks]{hyperref}

\usepackage{subcaption}
\usepackage{graphicx}
\usepackage[dvipsnames]{xcolor}
\usepackage{orcidlink}
\usepackage{multicol, multirow}
\usepackage{booktabs}
\usepackage{tikz}
\usepackage[edges]{forest}
\usepackage{amssymb, amsmath}
\usepackage{pifont}
\usepackage{cleveref}
\usepackage{siunitx}
\usepackage{pgfplots}
\pgfplotsset{compat=1.17} 
\usetikzlibrary{arrows.meta}
\usetikzlibrary{arrows}

\newcommand{\xmark}{\ding{55}}%
\newcommand{\cmark}{\ding{51}}%

\title{Grid-based Hybrid 3DMA GNSS and Terrestrial Positioning}

\author{
    Paul~Schwarzbach\orcidlink{0000-0002-1091-782X}, Albrecht~Michler\orcidlink{0000-0002-3434-3488}, Oliver~Michler\orcidlink{0000-0002-8599-5304}, \textit{TUD Dresden University of Technology}
    }

\begin{document}

\maketitle


\section*{BIOGRAPHY}

\textbf{Paul Schwarzbach} is a Research Assistant and Ph.D. student at the chair of Transport Systems Information Technology at TUD Dresden University of Technology since 2016. He received his diploma in traffic engineering also in 2016. His research interest focuses on probabilistic state estimation and tracking methods for intelligent transportation systems.
\\
\textbf{Albrecht Michler}  is a Research Assistant and Ph.D. student at the chair of Transport Systems Information Technology at TUD Dresden University of Technology since 2018. He received his diploma in traffic engineering also in 2018. His is interested in robust methods and integrity estimation for hybrid localization systems.
\\
\textbf{Prof. Oliver Michler} received his Diploma in Electrical Engineering from TUD Dresden University of Technology, Germany in 1993. From 1993 to 1997, he was a Research Assistant with the TUD and received his Ph.D. in 1999. Until 2008, he was a project manager at Video Audio Design GmbH, Research Associate at Fraunhofer Institute of Transportation and Infrastructure Systems and a Professor for Signal Processing and Electronic Measurement Techniques at University of Applied Sciences Dresden. Ever since, he became a full Professor for Transport Systems Information Technology at TUD, where he also became the Director of the Institute of Traffic Telematics in 2019.  In addition, he is a scientific advisory board member of several international conferences, organizations and start-up businesses. 

\section*{ABSTRACT}
The paper discusses the increasing use of hybridized sensor information for GNSS-based localization and navigation, including the use of 3D map-aided GNSS positioning and terrestrial systems based on different geometric measurement principles. However, both GNSS and terrestrial systems are subject to negative impacts from the propagation environment, which can violate the assumptions of conventionally applied parametric state estimators. Furthermore, dynamic parametric state estimation does not account for multi-modalities within the state space leading to an information loss within the prediction step. In addition, the synchronization of non-deterministic multi-rate measurement systems needs to be accounted.

In order to address these challenges, the paper proposes the use of a non-parametric filtering method, specifically a 3DMA multi-epoch Grid Filter, for the tight integration of GNSS and terrestrial signals. Specifically, the fusion of GNSS, Ultra-wide Band (UWB) and vehicle motion data is introduced based on a discrete state representation. Algorithmic challenges, including the use of different measurement models and time synchronization, are addressed. In order to evaluate the proposed method, real-world tests were conducted on an urban automotive testbed in both static and dynamic scenarios. 

We empirically show that we achieve sub-meter accuracy in the static scenario by averaging a positioning error of $\SI{0.64}{\meter}$, whereas in the dynamic scenario the average positioning error amounts to $\SI{1.62}{\meter}$.  

The paper provides a proof-of-concept of the introduced method and shows the feasibility of the inclusion of terrestrial signals in a 3DMA positioning framework in order to further enhance localization in GNSS-degraded environments. 


\section{INTRODUCTION}

The need for immersive localization systems based on a variety of technological solutions and their respective market potential is ever-growing \citep{euspa_lbs_requirements_2021}. Since stand-alone GNSS solutions typically do not meet the accompanying performance requirements, such as accuracy, availability and integrity, the use of additional sensor information is commonly applied \citep{GrejnerBrzezinska2016MultiSensorNavigationSystems}. Next to high-technology solutions including the incorporation of optical sensors for applications like automated driving, the use of map data and cooperative sensor information has greatly increased. This leads to a general hybridization of available augmentation inputs \citep{EgeaRoca_gnss_state_of_the_art_trends_2022}.

Due to the challenges for GNSS-based positioning in harsh urban environments, 3D map-aided (3DMA) GNSS positioning has received a lot of research attention in the past years \citep{groves_3dma_survey_part_i_2019}. Simultaneously, connected devices and corresponding infrastructure are rapidly emerging, enabling location-aware communication systems and a variety of location-based services. Based on different geometric measurement principles, such as time of arrival (ToA), angle of arrival (AoA) or time difference of arrival (TDoA), these terrestrial systems can also greatly benefit GNSS localization in dense or even GNSS-denied environments by applying hybrid or collaborative positioning \citep{Medina2020, Zhang_3dma_collaborative_positioning_2021}. 

Next to 3DMA GNSS, cooperative respectively collaborative positioning has been addressed in recent years \citep{Calatrava2023MassiveDifferencing}, also including the integration into smartphones \citep{Minetto_dgnss_cooperative_positioning_smartphone_2022}. By applying GNSS-only cooperative positioning, a correlation between observations can occur \citep{Zhang_study_multipath_spatial_correlation_gnss_collaborative_positioning_2021}, potentially hurting positioning performance.  However, since cooperative positioning is already based on the assumption of a radio connection between devices, the potential of using these radio signals for further augmentation arises. Examples include the fusion of GNSS with V2X DSRC radio \citep{Yan_tightly_coupled_gnss_dsrc_crlb_particle_filter_2022}, 5G \citep{Bai_gnss_5g_hybrid_multi_rate_measurements_2022} or UWB \citep{Huang_gnss_uwb_ppp_tightly_coupled_2022}. As previous studies have already discussed the benefits of hybrid GNSS and terrestrial positioning, e.g. by analyzing the geometric constellation \citep{Huang_GNSS_Collaborative_DOP_2016}, the conceptualization of collaborative localization frameworks already includes the idea of additional radio information, e.g. in \citep{Raviglione2022CollaborativeAwarenessCollaborativeInformation}. Furthermore, hybrid GNSS and terrestrial localization also enables the exploitation of demanding applications, such as seamless indoor/outdoor localization \citep{Bai_hybrid_indoor_outdoor_seamless_positioning_2022}.

The key contribution of the paper is expanding tight integration of GNSS and terrestrial signals based on a multi-epoch Grid Filter \citep{Schwarzbach2020TightIntegrationGNSSWSN}. The approach uses a dynamic model, which allows the incorporation of vehicle dynamics into the state estimation. The applied algorithmic toolchain will be detailed within the paper. In unison with GNSS, terrestrial systems are prone to the negative impacts of the propagation environment \citep{Schwarzbach_statistical_evaluation_2021}, including multipath and non-line-of-sight reception, leading to a violation of the assumptions of conventionally applied parametric state estimators, such as the Extended Kalman Filter (EKF). However, the usage of non-parametric filtering, such as the Grid Filter, allows a less stringent criteria formulation for the provided input data. Furthermore, parametric filtering approaches are prone to linearization errors as described in \citep{Julier_unscented_kalman_filter_2004}, which is further amplified in local terrestrial systems due to the geometric constellation of reference points and rovers.

\begin{figure}[htb!]
    \centering
    \begin{subfigure}[b]{0.33\textwidth}
    \centering
    \includegraphics[width=.8\linewidth]{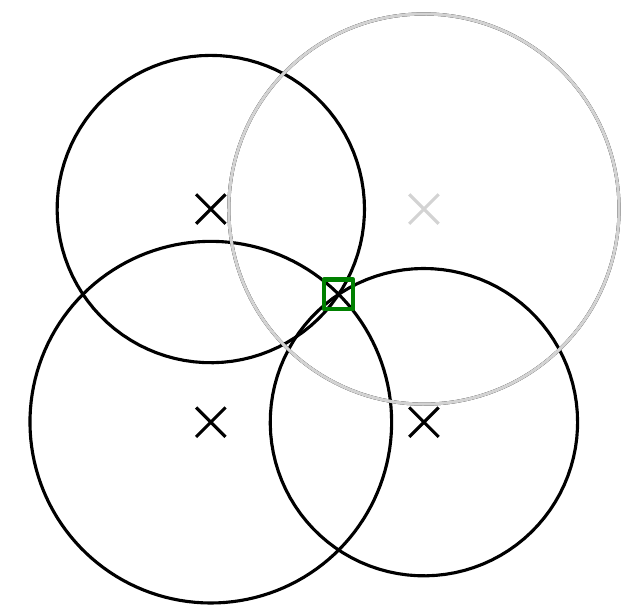}
    \caption{}
    \label{fig:radio_challenges_scenario}
    \end{subfigure}
    \centering
    \begin{subfigure}[b]{0.33\textwidth}
    \centering
    \includegraphics[width=1\linewidth]{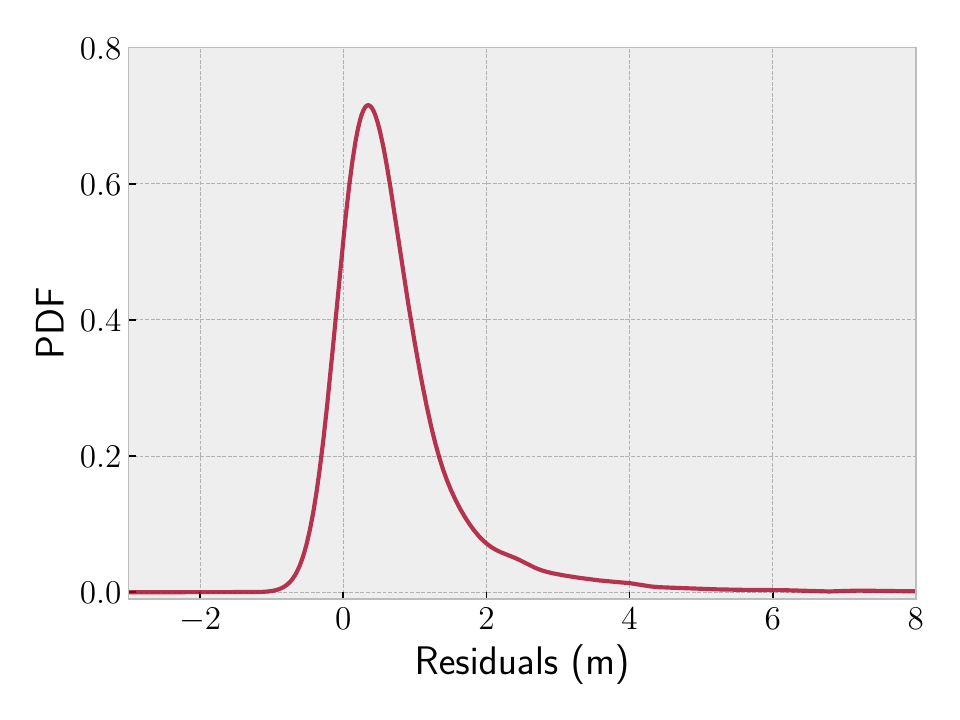}
    \caption{}
    \label{fig:radio_challenges_parametric}
    \end{subfigure}
    \centering
    \begin{subfigure}[b]{0.33\textwidth}
    \centering
    \includegraphics[width=.8\linewidth]{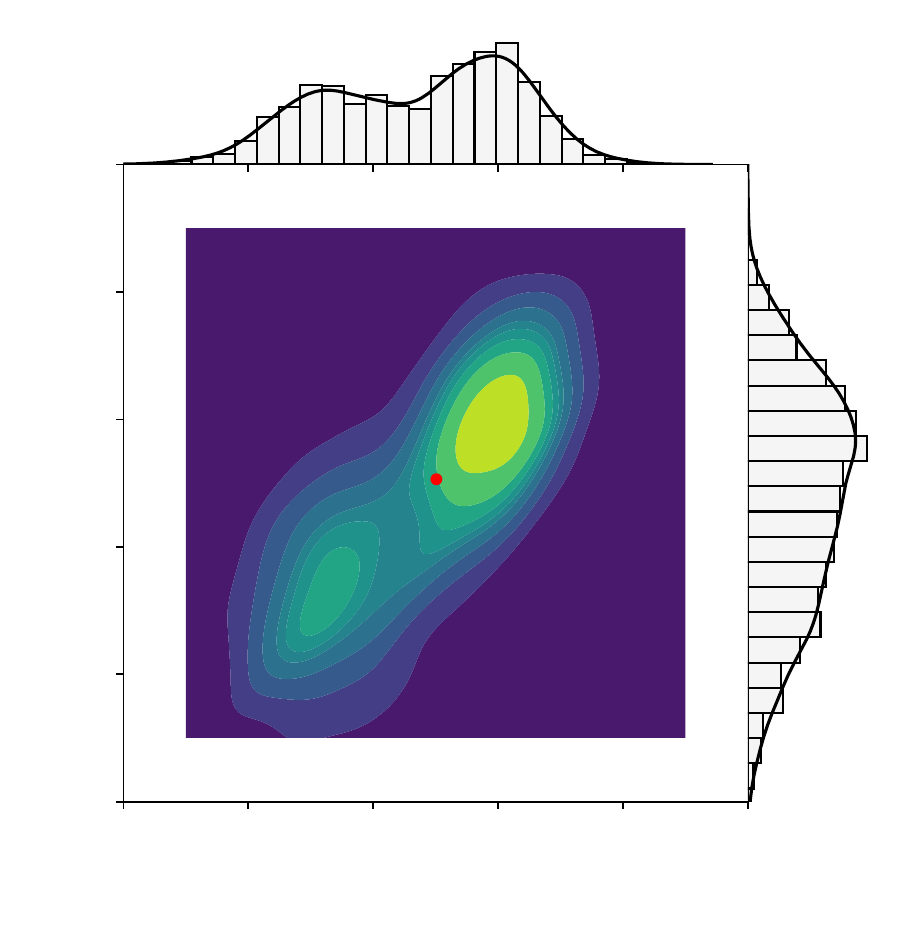}
    \caption{}
    \label{fig:radio_challenges_non_parametric}
    \end{subfigure}
    \caption{Problem formulation for NLOS reception in terrestrial ranging and corresponding challenges for state estimation: \textbf{(a)} Simulated ranging measurements (black) with 1 NLOS measurement (gray) and reference position (green); \textbf{(b)} Resulting right-skewed, non-gaussian ranging residuals corresponding to (a); \textbf{(c)} Resulting state-space multi-modalities and non-mitigated parametric estimation result (red).} 
    \label{fig:radio_challenges_states}
    \vspace{0.5cm}
\end{figure}

 Therefore, grid-based methods can handle both non-gaussian observation residuals and multi-modalities within the state space. This is emphasized in \cref{fig:radio_challenges_states}, which depicts the influence of measurement outliers (e.g. caused by NLOS reception) on both the positioning domain (right-skewed residual distribution) and the state space (multi-modality). In addition, the influence of the non-mitigated observation on parametric state estimation is shown. 

The paper presents the capabilities for a tight integration of heterogeneous terrestrial measurements, including ToA, AoA and TDoA observations, with the concept of 3DMA GNSS positioning \citep{Schwarzbach2020SingleDiff}. This is done by evolving a multi-epoch, 3DMA, grid-based Bayesian Filter, including an algorithmic generalization for the tightly coupled data fusion of the aforementioned geometric relations provided from terrestrial systems. Since the given data fusion problem formulation is reduced to a technology-independent integration of geometric relations based on spatial map data, a synergistic foundation for the future integration of opportunistic radio signals, for example including 5G-based TDoA or AoA observations or WiFi Fine Timing Measurements (FTM), is provided. 

In addition to addressing the measurement model and adapting it in accordance with hybrid GNSS and terrestrial Grid Filtering (selection of sampling probabilities and their parameter settings), we also present the integration of the motion step within the grid state space representation, allowing the incorporation of vehicle dynamics within the Grid Filter. This allows the propagation of a higher entropy of the estimated state distribution within the recursive filtering structure. Furthermore, time synchronization for tightly coupled integration is discussed, as a multi-rate sequential filtering approach is presented.

In addition to conceptual work and a formal description of the approach, a real-world study in both a dynamic and a static scenario is presented in order to empirically evaluate the proposed method. Here, a local Ultra-Wideband (UWB) real-time localization system is deployed in order to augment multi-constellation GNSS pseudorange observations. The study was conducted on a testbed for automated and connected driving in an urban environment in Germany. 

The rest of the paper is structured as follows: \cref{sec:theory} introduces the concept of the hybrid GNSS and terrestrial Grid Filter, including the discussion of potential radio technologies and respective measurement principles as well as the theory behind the hybrid Grid Filter and the integration of different geometric relations. In addition, a grid-based prediction step and a coping mechanism for multi-rate measurement synchronization is presented. The implementation of the provided theoretical background is detailed in \cref{sec:implementation}. Subsequently, \cref{sec:data} presents the conducted measurement campaign, including both a static and a dynamic scenario. Based on this, a detailed evaluation of the implemented method based on the surveyed data is provided. The paper concludes with a summary and an outlook for future research work.

\section{Hybrid GNSS and Terrestrial Positioning}
\label{sec:theory}

\subsection{Terrestrial Systems}
\label{ssec:terrestrial_systems}

\begin{figure}[!htbp]
    \centering
    \scalebox{.55}{\input{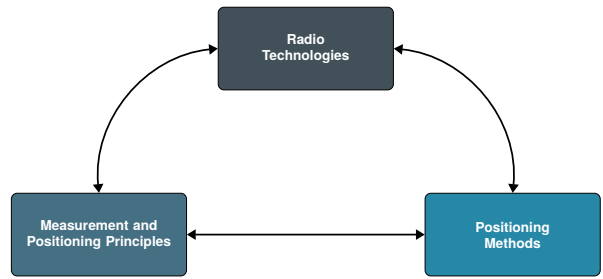}}
    \caption{Pillars of radio-based localization systems.}
    \label{fig:wireless_taxonomy}
\end{figure}

Radio-based and, more generally, wireless positioning is an interdisciplinary  engineering and scientific field. In the course of the years, different forms, interest groups and researchers have established themselves. However, there is no unified taxonomy for these systems in literature \citep{Pascacio_collaborative_indoor_positioning_2021}. This is due to the time difference of research and development works across different technologies. Nevertheless, it can be stated that a categorization of wireless localization systems is based on three pillars \citep{Esposito2011, Tariq2017}, which are presented in \cref{fig:wireless_taxonomy}.

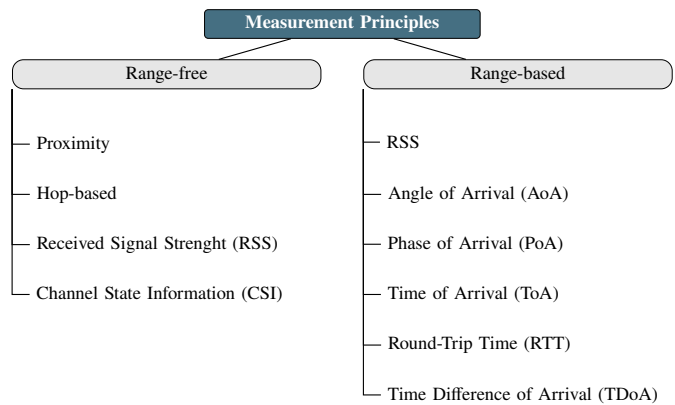
\begin{figure}[!htbp]
    \centering
     \scalebox{.7}{\usetikzlibrary{shadows,arrows.meta}

\tikzset{
    basic/.style  = {draw, text width=5cm, rectangle},
    parent/.style={basic, rounded corners=2pt, thin, align=center, fill=cyan!40!black, text=white},
    child/.style={basic, rounded corners=6pt, thin, align=center, fill=gray!20, text width=16em},
    grandchild/.style={basic, draw=none, text width=15em, align=left}
}

\begin{forest}
for tree={%
    thick,
    l sep=0.3cm,
    s sep=0.8cm,
    edge={semithick},
    where level=0{parent}{},
    where level=1{
        minimum height=0.2cm,
        child,
        parent anchor=south west,
        tier=p,
        l sep=0.05cm,
        for descendants={%
            grandchild,
            minimum height=0.2cm,
            anchor=120,
            edge path={
                \noexpand\path[\forestoption{edge}]
                (!to tier=p.parent anchor) |-(.child anchor)\forestoption{edge label};
            },
        }
    }{},
}
[\textbf{Measurement Principles}
    [Range-free
        [Proximity
            [Hop-based
                [Received Signal Strenght (RSS)
                    [Channel State Information (CSI)
                    ]
                ]
            ]
        ]
    ]
    [Range-based
        [RSS
            [Angle of Arrival (AoA)
                [Phase of Arrival (PoA)
                    [Time of Arrival (ToA)
                        [Round-Trip Time (RTT)
                            [Time Difference of Arrival (TDoA)
                            ]
                        ]
                    ]
                ]
            ]
        ]
    ]
]
\end{forest}}
    \caption{Classification of wireless measurement principles as a basis for position determination.}
    \label{fig:drahtlose_messprinzipien}
\end{figure}

In unison with GNSS localization, the basis for determining location-related variables is the derivation of signal properties, more precisely properties of the signal transmission, e.g., connectivity or physical quantities of the signal. In this stage, a distinction between range-free and range-based (also referred to as geometric) can be made. This is further visualized in \cref{fig:drahtlose_messprinzipien}, which also includes applicable measurement principles.

For geometry-free approaches, there is a strong (positive) correlation of the number of available infrastructure and end devices that participate in the positioning process. In general, the achievable positioning quality of these geometry-free approaches is comparatively low \citep{Chowdhury_advances_localization_wsn_survey_2016}. 

Range-based approaches on the other hand, allow the geometric interpretation of derived signal quantities. Applicable radio technologies and corresponding geometric measurement principles are mapped in \cref{tab:technology_mapping}.  This allows an overview on both system candidates and the discussion of advantages of the measurement principles, e.g. hardware availability, computational complexity or achievable accuracies.\footnote{ToA measurement principles are not applicable in terrestrial radio systems as a clock synchronization respectively a correction of the clock offsets between infrastructure and mobile devices is not given.}  Further information can be found in \citep{Zafari_IPS_Survey_2019, MendozaSilva_meta_review_ips_2019}.

\begin{table}[!htbp]
\centering
\caption{Mapping of range-based measurement principles and available technolgies for radio-based positioning.}
\begin{tabular}{@{}llllll@{}}
\toprule
     & WiFi & BLE & ZigBee & UWB & 5G \\ \midrule
RSS  & \cmark     &  \cmark            &   \cmark     & \cmark    & \cmark   \\
AoA  &  \cmark    &    \cmark          &   \cmark     & \cmark    &  \cmark  \\
PoA  &   \xmark   &      \cmark        &   \cmark     &  \xmark   &  \cmark  \\
RTT  &   \cmark   &    \xmark          &  \xmark      &  \cmark   &   \xmark \\
TDoA & \xmark     &    \xmark          &   \xmark     & \cmark    &  \cmark  \\ \bottomrule
\end{tabular}
\label{tab:technology_mapping}
\end{table}

In this context, promising system candidates for hybrid GNSS terrestrial positioning for transportation and logistic applications as well as seamless indoor outdoor positioning are given by:

\begin{itemize}
    \item WiFi-based Fine Timing Measurements (FTM) \citep{Yu2020WiFiFTM, Gentner_wifi_rtt_2020};
    \item BLE-based AoA implemented from version 5.1 \citep{Sambu_experimental_study_direction_finding_bluetooth_2022} and PoA Ranging \citep{Zand2019BLEPhaseRanging};
    \item 5G-based AoA and TDoA integration \citep{Talvitie_5g_train_tracking_tdoa_aoa_2019, Xhafa20215GAoATDoA};
    \item UWB-based Two-Way Ranging or TDoA \citep{Chiasson2023UWB_TDoA}.
\end{itemize}

As previously stated, the presented work focuses on the hybrid integration of terrestrial systems and GNSS based on a Grid Filter implementation. Therefore, individual quantities of different origin are abstracted to their geometric primitives and integrated on this level. By this tightly-coupled approach it is possible that no system-individual position solutions have to be available for an integration and the requirements concerning the available observation quantities can be reduced. Thus, individual observations can also be used to support the position information.

\subsection{Hybrid Grid Filter}
\label{ssec:hybrid_grid_filter}

The use of a grid-based representation of the state space for probabilistic filtering is fundamentally suitable for a variety of localization tasks and heterogeneous inputs \cite{thrun2005probabilistic}. In general, the proposed hybrid Grid Filter follows the well-studied prediction and observation structure of a Recursive Bayes Filter \citep{thrun2005probabilistic}, similar to the well-known Extended Kalman Filter or the Particle Filter. In this section we will present the general framework of a multi-epoch, hybrid Grid Filter and provide background information of identified challenges for hybrid Grid Filters: Generic measurement models, state propagation for dynamic applications and time synchronization for hybrid information fusion.

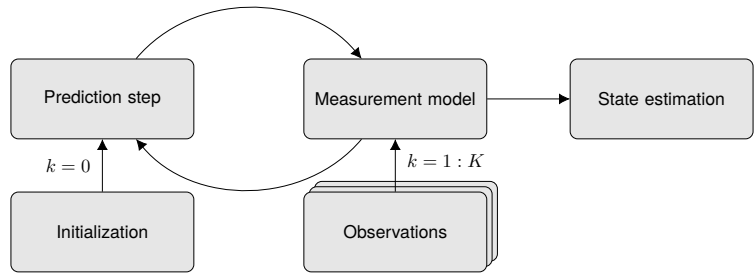
\begin{figure}[htbp!]
    \centering
    \resizebox{10cm}{!}{\tikzset{%
  >={Latex[width=2mm,length=2mm]},
            base/.style = {rectangle, rounded corners, draw=black,
                           minimum width=30mm, minimum height=1.5cm,
                           text width = 32mm, fill=black!10,
                           text centered},
  			standard/.style = {rectangle, rounded corners, draw=black,
                           minimum width=3cm, minimum height=1.5cm,
                           text width = 3.2cm, fill=black!10,
                           text centered},
            standard1/.style = {rectangle, rounded corners, draw=black,
                           minimum width=30mm, minimum height=1.5cm,
                           text width = 32mm, fill=black!10,
                           text centered},
       startstop/.style = {base, fill=red!30},
    activityRuns/.style = {base, fill=green!30},
         process/.style = {base, minimum width=2.5cm, fill=orange!15},
}

\begin{tikzpicture}[node distance=25mm,
    every node/.style={fill=white, font=\sffamily}, align=center]
  \node (start)             [standard]              {Initialization};
  \node (predict) [base, above of=start] {Prediction step};
  \node (observe) [base, right of=predict, xshift=3.0cm] {Measurement model};
  \node (data2) [standard, below of=observe, xshift=0.2cm, yshift=0.2cm] {Observations};
  \node (data1) [standard, below of=observe, xshift=0.1cm, yshift=0.1cm] {Observations};
  \node (data) [standard, below of=observe] {Observations};
  \node (estimate) [standard1, right of=observe, xshift=2.5cm] {State estimation};
    
   \draw[->] (start) -- node[midway,left, xshift=-0.1cm] {$k=0$} (predict);
   \draw[->] (data) -- node[midway,right, xshift=0.1cm, yshift=0.1cm] {$k=1:K$} (observe);
   \draw[->] (observe) -- (estimate);
    \draw[<-] (predict) to[bend right=50](observe);
    \draw[->] (predict) to[bend left=50](observe);
  \end{tikzpicture}}
    \caption{Generic RBF structure (according to \cite{Schwarzbach2020TightIntegrationGNSSWSN}).}
    \label{fig:rbf_structure}
\end{figure}

The general framework in accordance with the Recursive Bayes Filter (RBF) structure is given in \cref{fig:rbf_structure}. The following section will incorporate the accompanying calculation steps and their theoretical background as well as the final implementation of the method.

\subsubsection{Measurement Model}
\label{ssec:measurement}

As aforementioned, the focus of the work lies on a general integration of geometric inputs, complementary to global GNSS positioning. Similar to 3DMA likelihood-based localization frameworks, e.g. \citep{Groves2020LikelihoodRanging} or \citep{Ng2020Rangebased3DMA}, a translation of available measurements in a probabilistic manner needs to be performed. For simplification reasons, the following explanations concerning the implementation of the measurement models are detailed using a local two-dimensional and equidistant grid. A generalization using digital maps can easily be done by using discrete, multidimensional and globally referenced geodata, such as digital surface models (DSM) \citep{Schwarzbach2020SingleDiff}.  The basic computational steps are given in \cref{fig:grid_integration}:

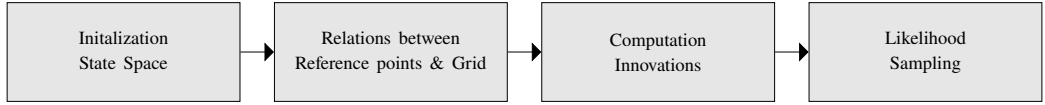
\begin{figure}[htbp!]
    \centering
    \resizebox{15cm}{!}{\usetikzlibrary{arrows.meta, mindmap, shadows}
\usetikzlibrary{arrows}
\usetikzlibrary{backgrounds}

\begin{tikzpicture}[node distance=5em,
    base/.style = {rectangle,  fill=black!10, draw=black,
                           minimum width=33mm, minimum height=15mm,
                           text width = 33mm,
                           text centered},
    frameless/.style = {rectangle, rounded corners, draw=white,
                           minimum width=30mm, minimum height=3mm,
                           text width = 30mm,
                           text centered},
    empty/.style = {rectangle, rounded corners, draw=black,
                           minimum width=5mm, minimum height=15mm,
                           text width = 27mm,
                           text centered}
    every node/.style={fill=white, font=\sffamily}, align=center]

  \node (init) [base] { \footnotesize Initalization \\ State Space};
  \node (relation) [base, right of=init, xshift=23mm] { \footnotesize Relations between\\ Reference points \& Grid};
  \node (innovation) [base, right of=relation, xshift=23mm] {\footnotesize Computation \\ Innovations};
  \node (sampling) [base, right of=innovation, xshift=23mm] {\footnotesize Likelihood \\ Sampling};
  
  \draw[black, -triangle 90] (init) -- (relation);
  \draw[black, -triangle 90] (relation) -- (innovation);
  \draw[black, -triangle 90] (innovation) -- (sampling);
  
  

  
  \end{tikzpicture}}
    \caption{Schematic approach to transform geometric relations into probabilistic grid-based state space.}
    \label{fig:grid_integration}
\end{figure}

First, the definition of a generic, discrete and finite state space according is required. The result of this initialization are $i = 1, \dots, I$ grid points $\boldsymbol{\mathrm{x}}_i$ of arbitrary dimension. Given $n = 1, \dots, N$ available reference points\footnote{Reference points can be given as global satellites, base stations of mobile radio, WiFi access points or sensor network anchors.}  $\boldsymbol{{x}}^nF = [x^n, y^n, z^n]^\intercal$, the geometric relations between the defined state space $\boldsymbol{\mathrm{x}}_i$ and $\boldsymbol{{x}}^n$ are considered by calculating the innovations $\boldsymbol{\mathrm{y}}_i^n$. This is done by calculating the difference of the present observations $\boldsymbol{\mathcal{Z}}^n$ and the known relations between grid points and reference points $\boldsymbol{\Gamma}_i^n$:

\begin{equation}
    \boldsymbol{\mathrm{y}}_i^n = \boldsymbol{\mathcal{Z}}^n - \boldsymbol{\Gamma}_i^n \; .
    \label{equ:innovations}
\end{equation}

The most common relations applicable for radio-based positioning are summarized in \cref{tab:innovation}.

\begin{table}[htbp!]
\renewcommand{\arraystretch}{1.5}
\centering
\caption{Calculation of geometric relations between reference and grid points.}
\label{tab:innovation}
\begin{tabular}{@{}cc@{}}
\toprule
 Geometrical relation & \multicolumn{1}{c}{Calculation} \\ \midrule
Distance (ToA / RTT)            & $\boldsymbol{\Gamma}_{\text{distance}}^{i,j} = \left | \left| \boldsymbol{\textrm{x}}^n - \boldsymbol{\textrm{x}}_i \right | \right |_2$                                                \\ \midrule
Hyperbolic (TDoA)  &  $\boldsymbol{\Gamma}_{\text{hyperbolic}}^{i,j} = \left | \left| \boldsymbol{\textrm{x}}^n - \boldsymbol{\textrm{x}}_i \right | \right |_2  - \left | \left| \boldsymbol{\textrm{x}}^{j+1} - \boldsymbol{\textrm{x}}_i \right | \right |_2$                                                 \\ \midrule
Angle (AoA)                &  $\boldsymbol{\Gamma}_{\text{angle}}^{i,j} = \arctan \frac{y^n - y_i}{x^n - x_i} $                                               \\ \bottomrule 
\end{tabular}
\end{table}

\begin{figure}[htbp!]
    \centering
    \begin{subfigure}[b]{0.27\textwidth}
    \centering
    \includegraphics[width=\linewidth]{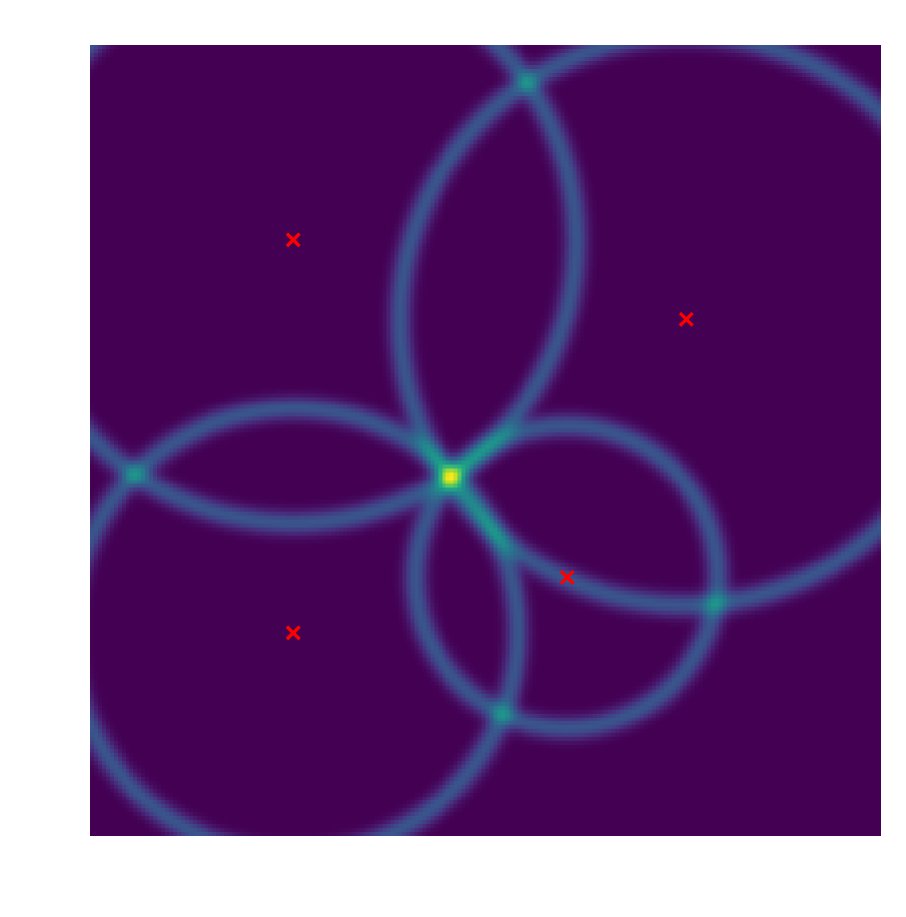}
    \caption{}
    \label{fig:radio_grid_tof}
    \end{subfigure}
    \centering
    \hspace{0.5cm}
    \begin{subfigure}[b]{0.27\textwidth}
    \centering
    \includegraphics[width=\linewidth]{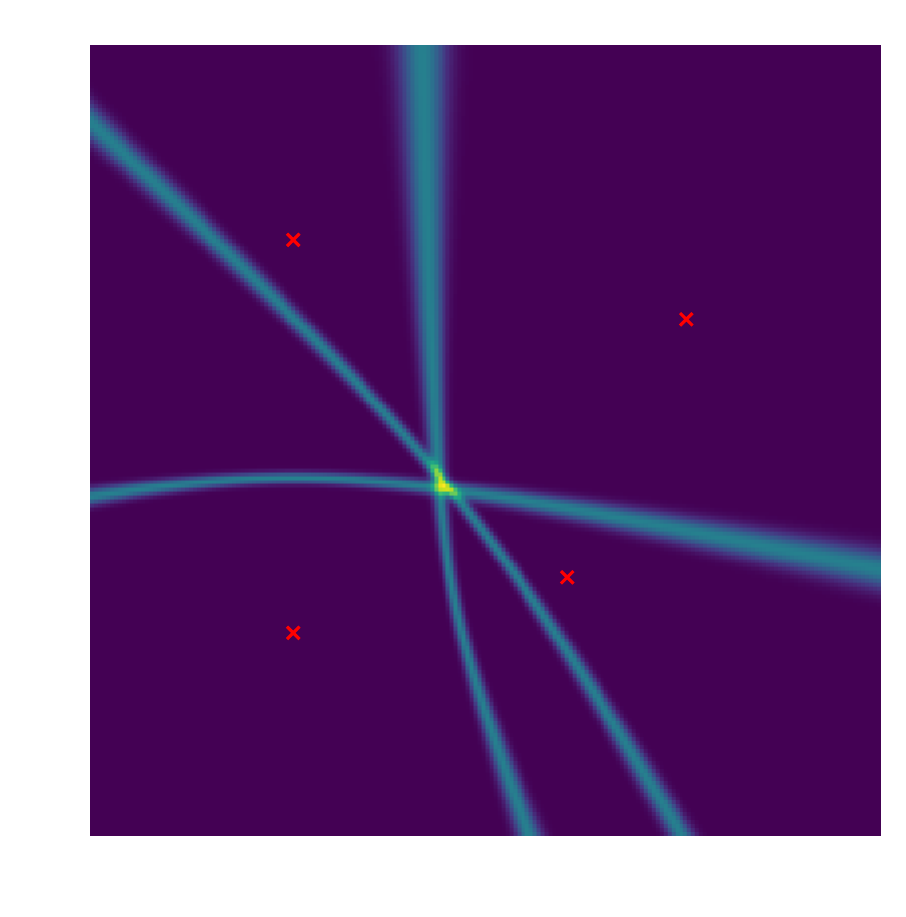}
    \caption{}
    \label{fig:radio_grid_tdoa}
    \end{subfigure}
    \hspace{0.5cm}
    \begin{subfigure}[b]{0.27\textwidth}
    \centering
    \includegraphics[width=\linewidth]{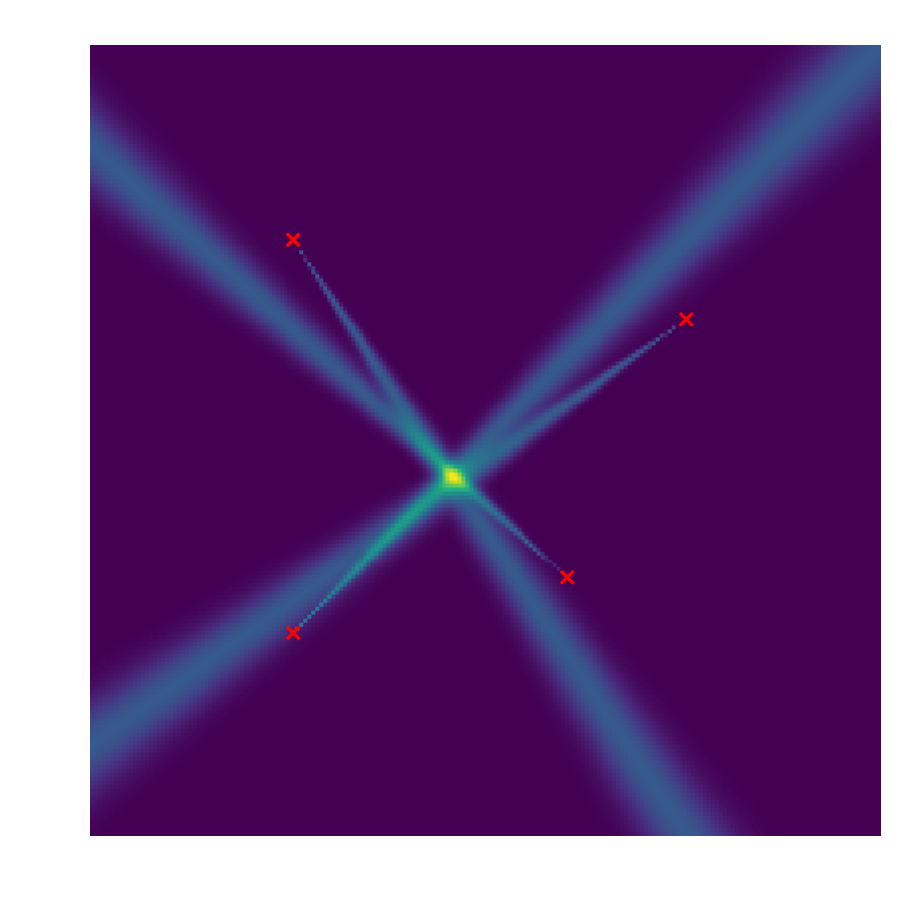}
    \caption{}
    \label{fig:radio_grid_aoa}
    \end{subfigure}
    \caption{Likelihood representation of geometric position lines in an equidistant two-dimensional grid based on obtaining spatial relations in a radio network (anchor points red, value of Likelihood color coded): (\textbf{a}) Distances, (\textbf{b}) Hyperbolic und (\textbf{c}) Angle.}
    \label{fig:radio_grid}
\end{figure}

For the innovations obtained in \cref{equ:innovations} for each grid point and each observation, a likelihood, is subsequently calculated under the assumption of a statistical model. This statistical model is available in the form of a probability density function (pdf) and characterizes the expected uncertainties of the respective observations. Given a generic statistical distribution (e.g. Gaussian) $\mathcal{D}(0, \boldsymbol{\Sigma})$, we can obtain the Likelihood of each position candidate from \cite{thrun2005probabilistic}:

\begin{equation}
    \textbf{p}_i^n(\boldsymbol{\mathcal{Z}}^n|\boldsymbol{\textrm{x}}_i) \leftarrow \mathcal{D}(\textbf{\textrm{y}}_i^{j}, \boldsymbol{\Sigma}) \; ,
\end{equation}

where $\boldsymbol{\Sigma}$ represents the scale parameter and therefore the statistical properties of the assumed pdf. In the case of Gaussian uncertainty, this represents the covariance matrix. The final combination of all observations respectively their likelihoods over all grid points is performed (here without consideration of the propagated likelihoods discussed in \cref{ssec:prediction}) by means of:

\begin{equation}
    \boldsymbol{p}_i = \eta \sum_j \textbf{p}_i^n(\boldsymbol{\mathcal{Z}}^n|\boldsymbol{\textrm{x}}_i) \; ,
    \label{equ:likelihood_prod}
\end{equation}

where $\eta$ represents the normalization factor according to Bayes' rule. The result of this calculation for the geometric relations given in \cref{tab:innovation} are depicted in \cref{fig:radio_grid}. Here, the Likelihood obtained from four reference points (red) is given. The statistical model applied is a Gaussian distribution.

In addition, Grid Filtering easily allows a combination of different geometrical relations within one or consecutive measurement steps, which also enables localization systems with less required infrastructure respectively reference points. The integration of GNSS and terrestrial observations is further detailed in \cref{sec:implementation}.

\subsubsection{Prediction step}
\label{ssec:prediction}

An important component of dynamic state estimation respectively multi-epoch filtering is the prediction step, which takes a motion model and/or sensor information about the object motion into account. For dynamic state estimation based on the EKF approach, velocities, direction of motion, accelerations, and others are considered in the course of mathematical modeling depending on the model degree \cite{Balzer_epe_ekf_gnss_2014}. For the estimation of these motion quantities, the use of EKF approaches and their derivatives has become established, since higher-dimensional state vectors can be implemented in a computationally efficient manner and these sensors satisfy the strict requirements of parametric estimation methods. This has also been applied in previous works addressing multi-epoch 3DMA filtering in \citep{Zhong2022MultiEpoch3DMA}.

Nevertheless, an integration of rudimentary motion information within the prediction step of the Grid Filter is possible and absolutely necessary in the context of the possible asynchronicity of observations of heterogeneous origin (cf. \cref{ssec:time_synchro}). The approach of the grid-based computation of the prediction step is based on the determination of the translation likelihoods between the realizations (position candidates) $\boldsymbol{\mathrm{x}}_i$ present in the state space. For this purpose, a probability-based plausibility check of the transition from one system candidate to another is performed based on dynamics data of an object or starting from model assumptions.

The two basic quantities for characterizing the motion of an object within a grid-based state representation are the object velocity $v_k$ and its heading $\theta_k$. The representation of the motion in this form is called odometry-based information \cite{thrun2005probabilistic}. The grid integration here is two-dimensional within the $x$-$y$ plane of the object to be located.

The dynamics information is processed individually within the likelihood grid and then combined. The underlying geometric relations are to be applied in analogy to the likelihood transfer of cooperative measured quantities listed in \cref{tab:innovation}. Here, a velocity measurement represents a time-scaled distance determination, where $\Delta T_k$ represent the temporal length of the prediction step as a function of the measurement rates and the availability of observation information. Furthermore, the heading represents an angular relation between the object orientation and a reference direction.

The peculiarity of the non-parametric estimation approach, compared to parametric estimators for dynamical systems (e.g. EKF), is that the motion prediction is performed not only for one position realization, in the case of parametric estimation this corresponds to the state vector consisting of the estimated mean values, but for all position hypotheses $\boldsymbol{\mathrm{x}}_i$. Thus, the motion prediction of non-parametric methods is associated with a comparatively significant increase in computational complexity. However, this also allows possibility to consider multi-modalities present in the state space. Thus, it is possible to obtain an increased information content within the state space in the prediction step compared to parametric approaches.

According to the definition of RBF, an existing velocity measurement or model assumption $v$ is also considered as a probabilistic quantity and thus represents in its simplest form an average value of the existing velocity in the direction of motion. This is further subject to an uncertainty $\sigma_{\text{v}}^2$, which can be parameterized empirically from available sensor data or model-based. The resulting prediction of motion is given as follows. First, for a candidate position $\boldsymbol{\mathrm{x}}_i$, the known distance $\boldsymbol{d}_{i,j}$ to all other candidate positions $\boldsymbol{\mathrm{x}}_j$ is calculated:

\begin{equation}
    \boldsymbol{d}_{i,j} = \boldsymbol{\Gamma}_{i,j} = \left | \left| \boldsymbol{\textrm{x}}_j - \boldsymbol{\textrm{x}}_i \right | \right |_2 \quad \text{with} \quad i \neq j \;.
    \label{equ:grid_predict_distance}
\end{equation}

Subsequently, the determination of the residuals $\boldsymbol{\mathrm{y}}_{i,j}^{\text{v}}$ between the determined distances $\boldsymbol{d}_{i,j}$ and the present velocity measurement $v_k$ in combination with the time difference $\Delta T_k$ is performed:

\begin{equation}
    \boldsymbol{\mathrm{y}}_{i,j}^{\text{v}} = v_k \cdot \Delta T_k - \boldsymbol{d}_{i,j} \; .
\end{equation}

Thus, concentric residuals arise, which have their origin in the used position hypothesis and whose radius corresponds to the time-scaled distance equivalent of the velocity measurement. In analogy to the calculation of the likelihood of the observations, a probabilistic sampling of the residuals based on an assumed stochastic model is performed to determine the predicted likelihood $\boldsymbol{\overline{p}}_{i,j}^{\text{v}}$, which is assumed to follow a Gaussian Distribution $\mathcal{N}(0, \sigma_v^2)$ :

\begin{equation}
    \boldsymbol{\overline{p}}_{i,j}^{\text{v}} \leftarrow \mathcal{N}(\boldsymbol{\textrm{y}}_{i,j}^{\text{h}}, \sigma_{\text{v}}^2) \;.
\end{equation}

The given calculation steps are repeated for all  grid points. As a final step a calculation of the total likelihood:

\begin{equation}
    \overline{\boldsymbol{p}}_i^{\text{v}} = \sum_j \overline{\boldsymbol{p}}_{i,j}^{\text{v}} \; .
    \label{equ:grid_predict_sum}
\end{equation}

The integration of the \textit{heading} information is done in analogy considering the measured or assumed \textit{heading} $\theta_k$, its uncertainty $\sigma_2^\text{h}$, the residuals of the \textit{heading} $\boldsymbol{\mathrm{y}}_{i,j}^{\text{h}}$ and the predicted likelihood $\overline{\boldsymbol{p}}_i^{\text{h}}$ by means of:

\begin{align}
    \boldsymbol{\alpha}_{i,j} &= \arctan \frac{y_j - y_i}{x_j - x_i} \quad \text{mit} \quad i \neq j \;, \\
    \boldsymbol{\mathrm{y}}_{i,j}^{\text{h}} &= \theta_k - \boldsymbol{\alpha}_{i,j}\;, \\
    \boldsymbol{\overline{p}}_{i,j}^{\text{h}} &\leftarrow \mathcal{N}(\boldsymbol{\textrm{y}}_{i,j}^{\text{h}}, \sigma_{\text{h}}^2) \;, \\
    \overline{\boldsymbol{p}}_i^{\text{h}} &= \sum_j \overline{\boldsymbol{p}}_{i,j}^{\text{h}} \; .
\end{align}

Given the availability of both measurements $v_k$ and $\theta_k$, the resulting likelihood of the prediction step is obtained taking into account the normalized likelihood of the observation model (cf. \cref{equ:likelihood_prod}) of the last measurement epoch:

\begin{equation}
    \overline{\boldsymbol{p}}_i = \sum_i \left (\overline{\boldsymbol{p}}_i^{\text{v}} \cdot \overline{\boldsymbol{p}}_i^{\text{h}} \right ) \; \boldsymbol{p}_{i} \; .
    \label{equ:grid_predict_speed_heading}
\end{equation}

\begin{figure}[htbp!]
    \centering
    \begin{subfigure}[b]{0.27\textwidth}
    \centering
    \includegraphics[width=\linewidth]{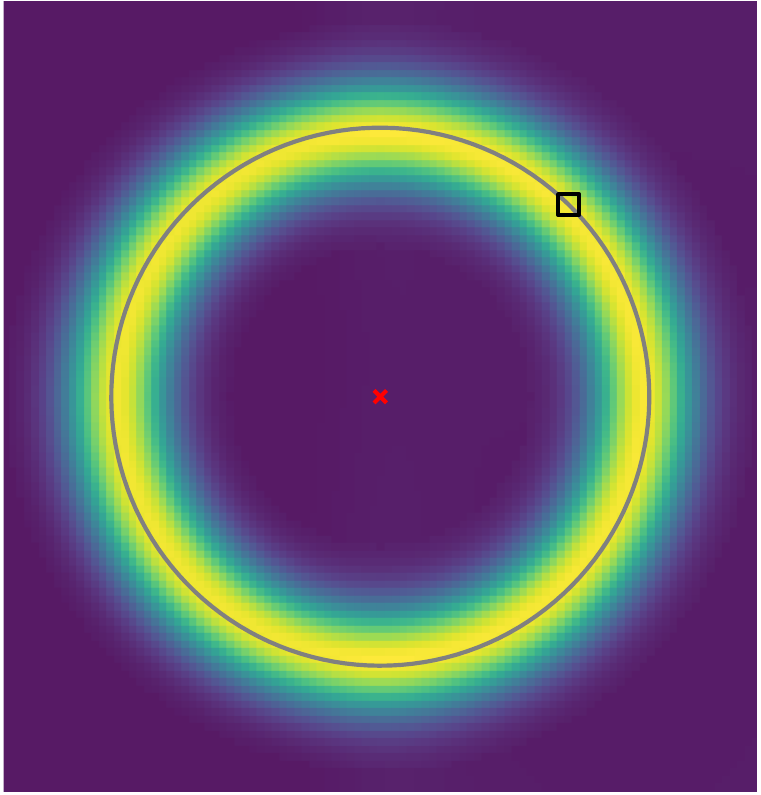}
    \caption{}
    \label{fig:grid_predict_speed}
    \end{subfigure}
    \hspace{0.5cm}
    \centering
    \begin{subfigure}[b]{0.27\textwidth}
    \centering
    \includegraphics[width=\linewidth]{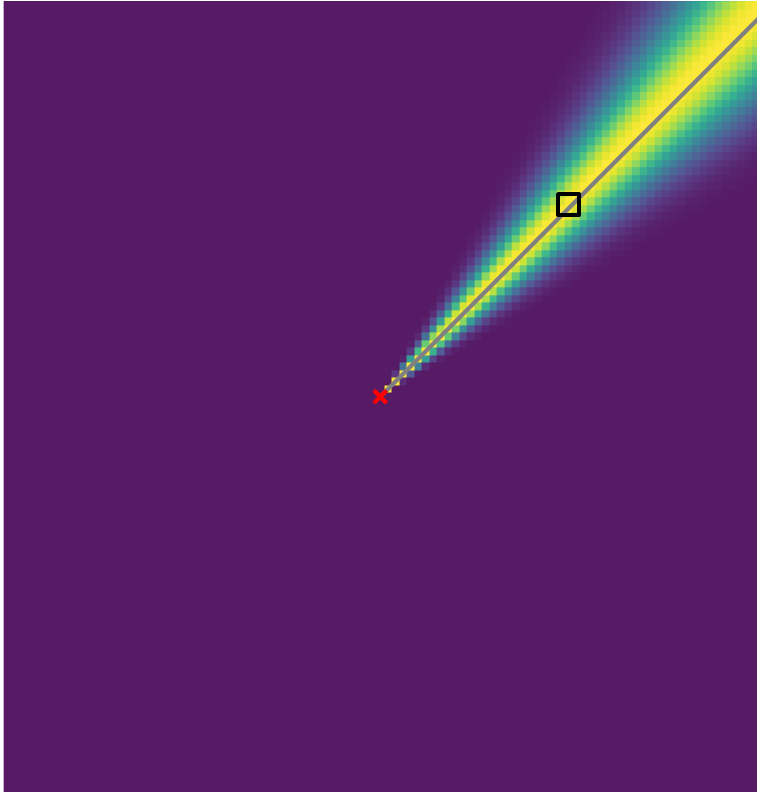}
    \caption{}
    \label{fig:grid_predict_heading}
    \end{subfigure}
    \hspace{0.5cm}
    \centering
    \begin{subfigure}[b]{0.28205\textwidth}
    \centering
    \includegraphics[width=\linewidth]{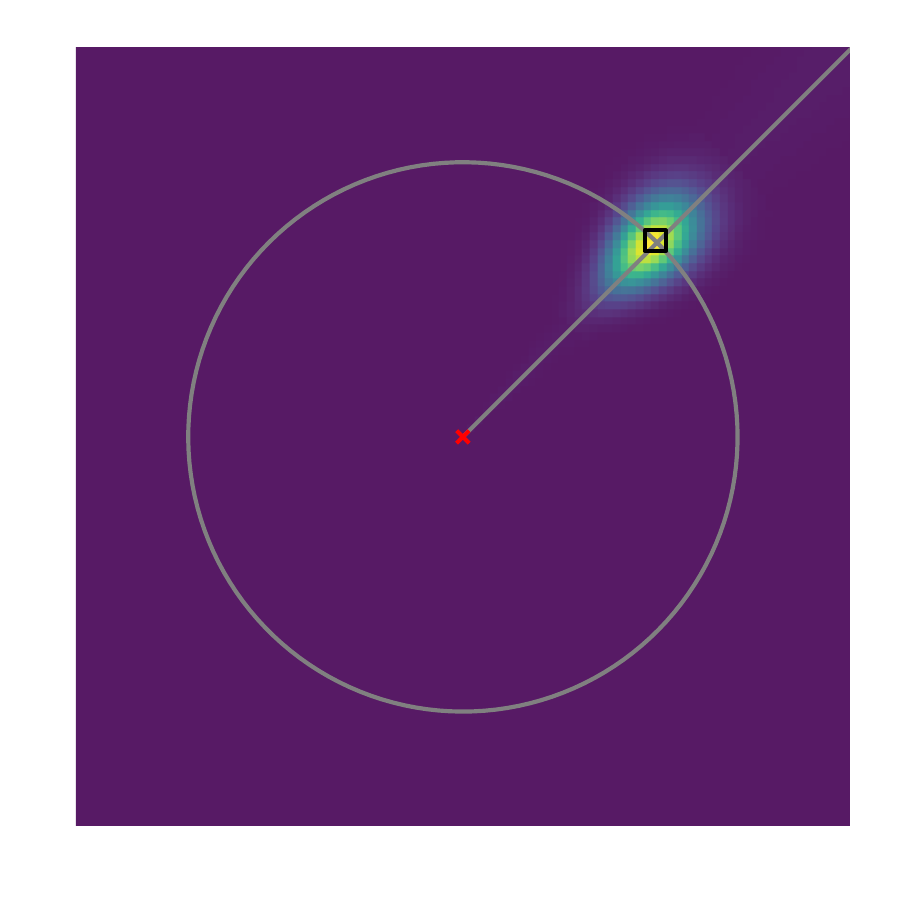}
    \caption{}
    \label{fig:grid_predict_fusion}
    \end{subfigure}
    \caption{Visualization of the grid-based prediction step based on dynamics information or model assumptions (true values are given in grey, red indicates the position candite of interest $i$): \textbf{(a)} Velocity likelihood $\overline{\boldsymbol{p}}_i^{\text{v}}$ $v_k$; \textbf{(b)} Heading likelihood $\overline{\boldsymbol{p}}_i^{\text{h}}$ $\theta_k$; \textbf{(c)} Fusion according to \cref{equ:grid_predict_speed_heading}.}
    \label{fig:grid_predict}
\end{figure}

A visualization of the calculation step for a candidate position within a probability grid is shown as an example in \cref{fig:grid_predict}.

Due to the finite nature of the discrete state space representation, a system propagation for global positioning problems needs to also be accounted, as moving objects are able to leave the initially defined state space. A procedure for this task is detailed in \citep{Zhong2022MultiEpoch3DMA} and therefore is not further addressed in this paper.

\subsubsection{Time Synchronization}
\label{ssec:time_synchro}

Time synchronization for hybrid positioning methods, especially with multi-rate systems has recently been addressed for GNSS and 5G \cite{Bai_gnss_5g_hybrid_multi_rate_measurements_2022} or GNSSS and UWB in \cite{Guo2023ekf_time_calibration}. Thereby, a possible synchronicity respectively handling of asynchronous observations especially considering different measurement rates is imperative, as this can lead to cumulative errors in the data fusion process. This concern is further enhanced in the presence of additional terrestrial systems due to the unknown time-offset of the individual systems and unequal, potentially non-deterministic update rates. 

As pointed out in \citep{Guo2023ekf_time_calibration} it is essential, that observations at different rates need to be processed. In \citep{Retscher2023FusionGNSSandUWBLeastSquares}, individual measurement models for GNSS and combined GNSS/UWB are defined. However, a simultaneous processing can only be realized if the time offset between the systems is compensated, especially in dynamic scenarios. Another major challenge poses the handling of out of sequence measurements \citep{Muntzinger2010OutOfSequenceMeasurements}. However, since we only discuss the integration of radio based inputs, the present measurement rates are comparably low. Therefore, only a timing consistency check between the time stamps of the sensor information is performed. In addition, we opt to implement a multi-rate sequential filtering approach based on the sensor time stamps of the individual sensor systems. 

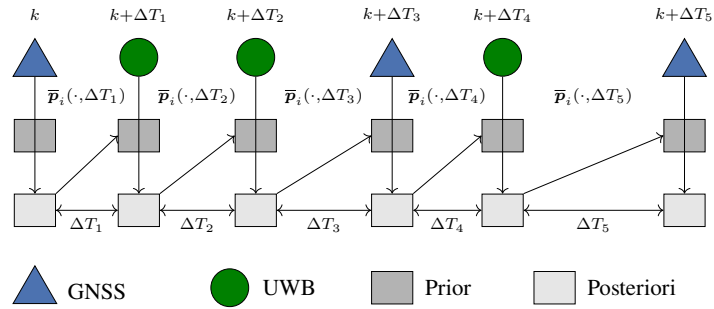
\begin{figure}[htbp!]
    \centering
    \resizebox{.55\textwidth}{!}{    \definecolor{blue}{RGB}{76,114,176}
    \definecolor{green}{RGB}{0,128,0}
    
    \begin{tikzpicture}[node distance=5em,
    gnss/.style =           {regular polygon,
                            regular polygon sides=3,fill=blue, draw=black, 
                            minimum size=0.01mm,
                            text width=.4mm,
                           },
    uwb/.style =          {circle, fill=green, draw=black,
                           minimum width=5mm,
                           text width = 3mm,
                           },
    state/.style = {rectangle, draw=black, fill=black!10!white,
                           minimum width=5mm, minimum height=5mm,
                           text width = 4mm,
                           text centered},
    prior/.style = {rectangle, draw=black, fill=black!30!white,
                           minimum width=5mm, minimum height=5mm,
                           text width = 4mm,
                           text centered},
    empty/.style = {rectangle, rounded corners, draw=black,
                           minimum width=5mm, minimum height=5mm,
                           text width = 4mm, text height=4mm,
                           text centered}
    every node/.style={fill=white, font=\sffamily}, align=center]


    \node (gnss1) [gnss] at (0.5, 2) {};
    \node (gnss2) [gnss] at (6., 2) {};
    \node (gnss3) [gnss] at (10.5, 2) {};
    \node (uwb1) [uwb] at (2.1, 2.1) {};
    \node (uwb2) [uwb] at (3.9, 2.1) {};
    \node (uwb3) [uwb] at (7.7, 2.1) {};

    \node (state1) [state, below of=gnss1, yshift=-5mm] {};
    \node (state2) [state, below of=uwb1, yshift=-6mm] {};
    \node (state3) [state, below of=uwb2, yshift=-6mm] {};
    \node (state4) [state, below of=gnss2, yshift=-5mm] {};
    \node (state5) [state, below of=uwb3, yshift=-6mm] {};
    \node (state6) [state, below of=gnss3, yshift=-5mm] {};

    \node (prior1) [prior, below of=gnss1, yshift=6.5mm] {};
    \node (prior2) [prior, below of=uwb1, yshift=5.5mm] {};
    \node (prior3) [prior, below of=uwb2, yshift=5.5mm] {};
    \node (prior4) [prior, below of=gnss2, yshift=6.5mm] {};
    \node (prior5) [prior, below of=uwb3, yshift=5.5mm] {};
    \node (prior6) [prior, below of=gnss3, yshift=6.5mm] {};

    \draw[->] (gnss1.south) -- (state1.north);
    \draw[->] (uwb1.south) -- (state2.north);
    \draw[->] (uwb2.south) -- (state3.north);
    \draw[->] (gnss2.south) -- (state4.north);
    \draw[->] (uwb3.south) -- (state5.north);
    \draw[->] (gnss3.south) -- (state6.north);

    \node (x1) [text width=3mm, above of= gnss1, yshift=-10mm] {$\scriptstyle k$}; 
    \node (x2) [text width=12mm, above of= uwb1, yshift=-11.1mm] {$\scriptstyle k+\Delta T_1$};
    \node (x3) [text width=12mm, above of= uwb2, yshift=-11.1mm] {$\scriptstyle k+\Delta T_2$};
    \node (x4) [text width=12mm, above of= gnss2, yshift=-10mm] {$\scriptstyle k+\Delta T_3$};
    \node (x5) [text width=12mm, above of= uwb3, yshift=-11.1mm] {$\scriptstyle k+\Delta T_4$};
    \node (x6) [text width=12mm, above of= gnss3, yshift=-10mm] {$\scriptstyle k+\Delta T_5$};

    \draw[->] (state1.north east) to node [below, yshift=13mm] (s1) { $ \scriptstyle\overline{\boldsymbol{p}}_i(\cdot, \Delta T_1)$} (prior2.west);

    \draw[->] (state2.north east) to node [below, yshift=13mm] (s2) { $\scriptstyle \overline{\boldsymbol{p}}_i(\cdot, \Delta T_2)$} (prior3.west);

    \draw[->] (state3.north east) to node [below, yshift=13mm] (s3) { $\scriptstyle \overline{\boldsymbol{p}}_i(\cdot, \Delta T_3)$} (prior4.west);

    \draw[->] (state4.north east) to node [below, yshift=13mm] (s4) { $\scriptstyle \overline{\boldsymbol{p}}_i(\cdot, \Delta T_4)$} (prior5.west);

    \draw[->] (state5.north east) to node [below, yshift=13mm] (s5) { $\scriptstyle \overline{\boldsymbol{p}}_i(\cdot, \Delta T_5)$} (prior6.west);
    
    \draw[<->] (state1.east) to node [below] (t1) {$\scriptstyle \Delta T_1$} (state2.west);   
    \draw[<->] (state2.east) to node [below] (t2) {$\scriptstyle \Delta T_2$} (state3.west); 
    \draw[<->] (state3.east) to node [below] (t3) {$\scriptstyle \Delta T_3$} (state4.west); 
    \draw[<->] (state4.east) to node [below] (t4) {$\scriptstyle \Delta T_4$} (state5.west); 
    \draw[<->] (state5.east) to node [below] (t5) {$\scriptstyle \Delta T_5$} (state6.west); 

    \node (leg_sym1) [gnss] at (0.5, -1.5) {};
    \node (leg_tex1) [text width=5mm, right of=leg_sym1, xshift=-10mm] {GNSS}; 

    \node (leg_sym2) [uwb] at (3.5, -1.45) {};
    \node (leg_tex2) [text width=5mm, right of=leg_sym2, xshift=-10mm] {UWB};

    \node (leg_sym3) [prior] at (6, -1.45) {};
    \node (leg_tex3) [text width=5mm, right of=leg_sym3, xshift=-10mm] {Prior};
    
    \node (leg_sym4) [state] at (8.5, -1.45) {};
    \node (leg_tex4) [text width=5mm, right of=leg_sym4, xshift=-10mm] {Posteriori};
    
\end{tikzpicture}}
    \caption{Schematic representation of fusion of multirate measurements as sequential data fusion based on the presented prediction step.}
    \label{fig:grid_zeitsynchro}
\end{figure}

Essentially, the multi-rate measurements are compensated by the introduced motion model, accounting for possible pose changes of the object within a measurement update window. This procedure is schematically presented in \cref{fig:grid_zeitsynchro}. Since, a grid-based motion step is implemented (cf.~\cref{ssec:prediction}) a high entropy about the systems state is stored within the state space. Another advantage of the given approach is that once a position fix is achieved, there is generally no limitation on the amount of available measurements to perform the described measurement step. This enables the utilization of system-individual observations at different measurement epochs, while also allowing the accounting of potential multi-modalities.

\subsubsection{State estimation}\label{ssec:state_estimation}

Given the calculation of an overall Likelihood consisting of the prediction step and hybrid measurement models, the final state estimation has to be performed in order to obtain the current receiver position. This can either be achieved by performing a maximum likelihood respectively for the presented Bayesian approach a maximum a-posteriori (MAP) estimation \citep{BarShalom_2001_Estimation} or by performing a weighted mean (WM) given all samples \citep{Minetto_phdtesis_gnss_only_collaborative_positioning_2020}. However, both approaches pose a variety of challenges for global positioning as summarized in \cref{tab:state_estimation_pros_cons}.

\begin{table}[htbp!]
\centering
\caption{Advantages and disadvantages of conventional state estimators for grid-based estimation.}
\begin{tabular}{@{}p{1.cm}p{6.75cm}p{7.25cm}@{}}
\toprule
                         & \multicolumn{1}{c}{Advantages}                                                       & Disadvantages                                                                                                                                    \\ \midrule
\multirow{2}{*}{{MAP}}          & Maximum in case of multimodal distribution in state space & accuracy of estimation correlated with grid resolution and can only correspond to defined position hypotheses \\
                                        & Simple, efficient implementation                           & Potential loss of information, since spatial realization of likelihood is not taken into account.             \\ \midrule
\multirow{1}{*}{WM}  &  Total information of the likelihood of the state space is considered                                               &      State space covers a large area for global positioning, therefore irrelevant areas are considered.                                                                                                                                     \\ \bottomrule
\end{tabular}
\label{tab:state_estimation_pros_cons}

\end{table}

Therefore, similar to we apply a two-step approach for deriving the current state estimation $\hat{\boldsymbol{\mathcal{X}}}_k$ at time step $k$:

\begin{enumerate}
    \item Calculation of MAP by maximizing the obtained likelihood function
    \item Calculation of WM using the MAP as circle center given an user-defined radius $\mathrm{R}$ including $I_{\mathrm{R}}$ grid points.
\end{enumerate}

This yields the state estimation $\hat{\boldsymbol{\mathrm{x}}}_k = \left [ \hat{\boldsymbol{x}}_k, \hat{\boldsymbol{y}}_k, \hat{\boldsymbol{z}}_k \right ]^{\intercal}$ with: 
    
\begin{equation}
    \hat{\boldsymbol{x}}_k = \frac{\sum_i^{I_{\mathrm{R}}} \boldsymbol{p}_i(x_i| \boldsymbol{\mathcal{Z}}) \cdot \boldsymbol{{x}}_i}{\sum_i^{I_{\mathrm{R}}} \boldsymbol{p}_i(x_i| \boldsymbol{\mathcal{Z}})} \quad 
   \hat{\boldsymbol{y}}_k =\frac{\sum_i^{I_{\mathrm{R}}} \boldsymbol{p}_i({y}_i| \boldsymbol{\mathcal{Z}}) \cdot \boldsymbol{{y}}_i}{\sum_i^{I_{\mathrm{R}}} \boldsymbol{p}_i({y}_i| \boldsymbol{\mathcal{Z}})} \quad 
   \hat{\boldsymbol{z}}_k = \frac{\sum_i^{I_{\mathrm{R}}} \boldsymbol{p}_i({z}_i| \boldsymbol{\mathcal{Z}}) \cdot \boldsymbol{{z}}_i}{\sum_i^{I_{\mathrm{R}}} \boldsymbol{p}_i({z}_i| \boldsymbol{\mathcal{Z}})} \; ,
   \label{equ:state_estimation_weighted_mean}
\end{equation}

\section{Implementation}
\label{sec:implementation}
\subsection{Overview}

\begin{figure}[htb!]
    \centering
    \includegraphics[width=.77\linewidth]{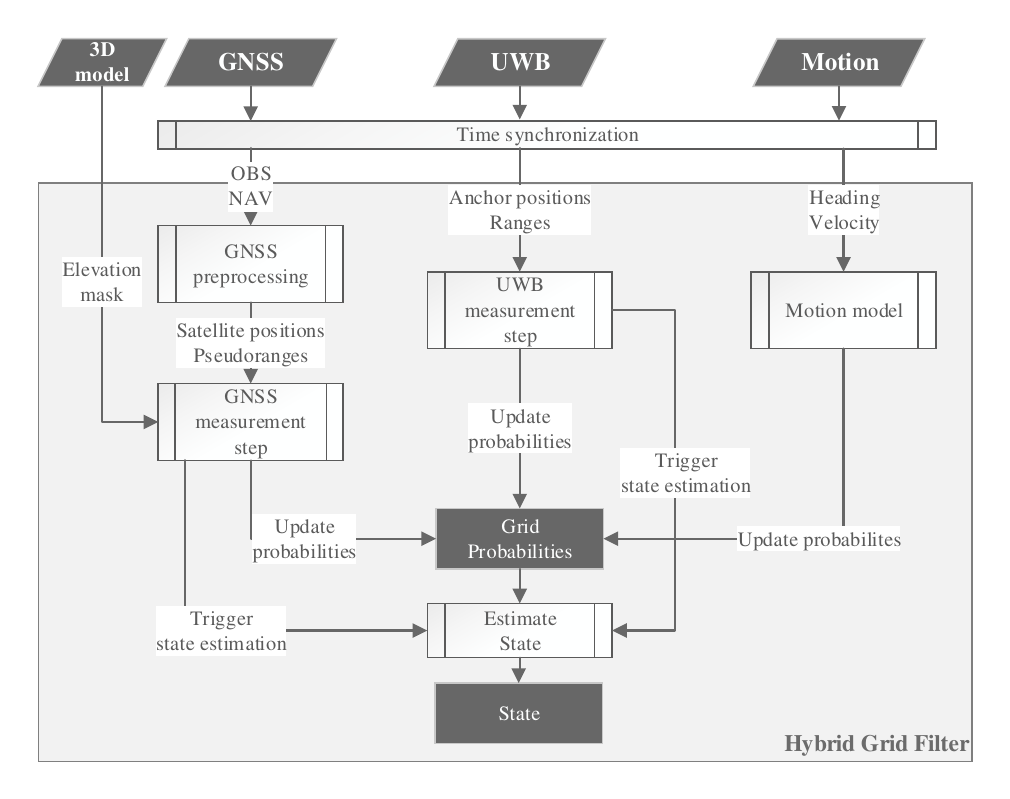}
    \caption{Flow chart of the implementation of a hybrid GNSS-UWB-motion Grid Filter.}
    \label{fig:pap}
\end{figure}

The implementation of the approach is based on the theoretical models and design considerations outlined in \cref{sec:theory}. An overview is provided in \cref{fig:pap}. The sensor components are synchronized to match a common time frame. As for GNSS, sensor data consists of satellite ephemeris data (NAV) and estimates of pseudoranges, carrier-phase, doppler and carrier-to-noise-ratio (OBS). In a preprocessing step, satellite positions are calculated and pseudoranges are corrected for error components, which can be modeled. This includes satellite clock bias and atmospheric errors as described in basic GNSS literature \citep{Kaplan2005}. The GNSS measurement step estimates the likelihood of each grid point based on given satellite positions and corrected pseudoranges. The measurement model is described in detail in \cref{ssec:gnss_meas}. The respective UWB measurement model relies on given anchor positions and ranging measurements for likelihood estimation. Furthermore, a motion model is incorporated, which uses heading and velocity measurements for likelihood estimation. Those measurements can be based on IMU sensor data, odometry or other sensor hardware. In case of absence of these input information, generic motion models can be applied. The resulting likelihood of each measurement step is then used to update the grid probabilities. Furthermore, each GNSS or UWB step triggers the state estimation. With this, the state vector is updated based on the current grid probability space as discussed in \cref{ssec:state_estimation}.

\subsection{Implementation of measurement models}

\begin{figure}[htbp!]
    \centering
    \begin{subfigure}[b]{0.45\textwidth}
    \centering
    \includegraphics[width=\linewidth]{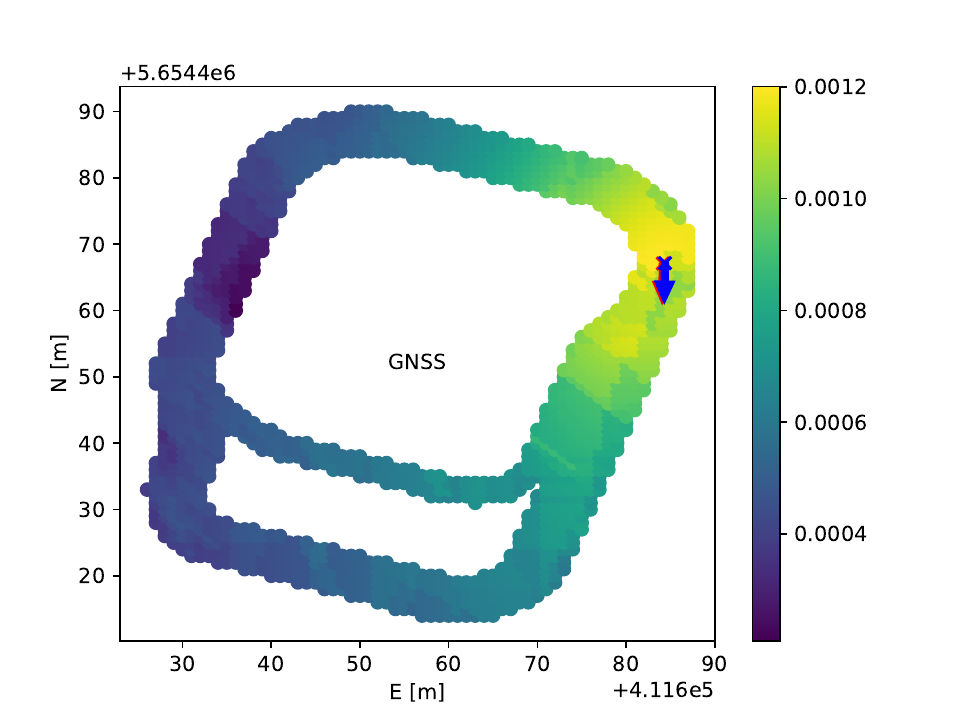}
    \caption{GNSS}
    \label{fig:gnss}
    \end{subfigure}
    \centering
    \begin{subfigure}[b]{0.45\textwidth}
    \centering
    \includegraphics[width=\linewidth]{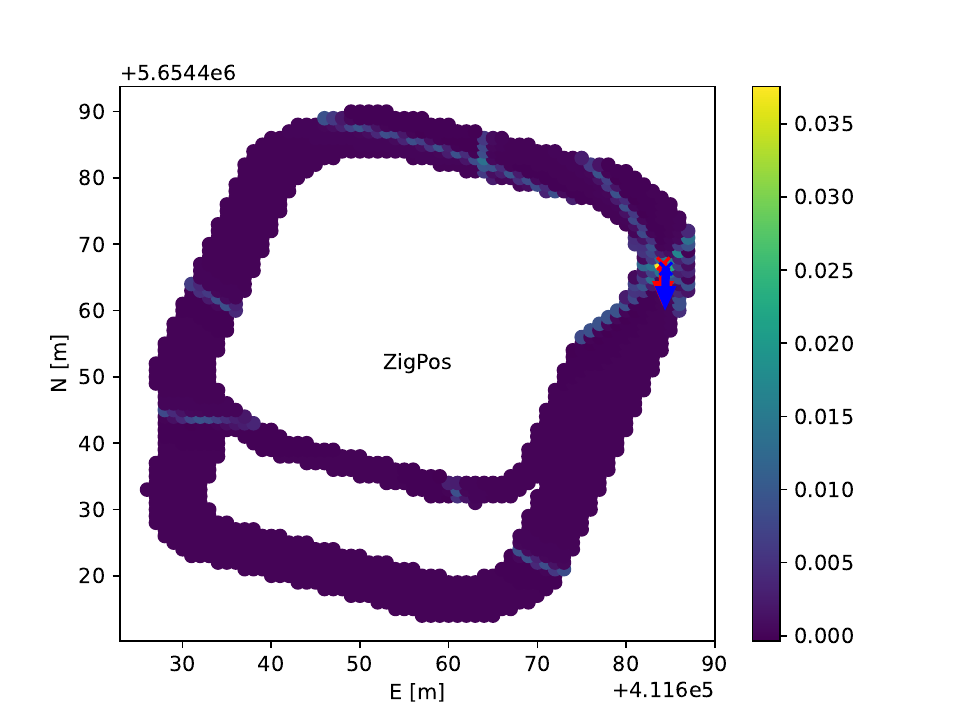}
    \caption{Terrestrial}
    \label{fig:zp}
    \end{subfigure}
    \caption{PDF after the measurement step. Reference vector (blue) and current estimate (red). Calculated Likelihoods are color coded.}
    \label{fig:measurement_models}
\end{figure}

The section details the implementation of the GNSS and terrestrial measurement model. Realizations of these individual Likelihoods are visualized in \cref{fig:measurement_models}. The figures are obtained from the surveyed dataset described in \cref{sec:data}. For both figures, the respective priors are ignored. The current state estimation based on the observation step and the corresponding reference position are also indicated.

\subsubsection{GNSS measurement model}\label{ssec:gnss_meas}

The presented work utilizes the well-studied Between Satellite Single Differencing \citep{groves_3dma_survey_part_i_2019, Suzuki_3dma_particle_filter_journal_2019, Schwarzbach2020SingleDiff}, suitable for code-based 3DMA positioning approaches using discrete state space representations. This is due to the removal of the receiver clock error, whose discretization is not feasible. 

The generic model for a pseudorange $\rho_r^s$  is given as \citep{Langley_springer_handbook_gnss_introduction_2017}:

\begin{equation}
 \rho_r^s = d_r^s + c \cdot (\delta t_r - \delta t^s) + \delta_{\text{ion}}^s + \delta_{\text{tro}}^s + \varepsilon_r^s \;,
 \label{equ:pseudorange_model}
\end{equation}

where $d_r^s = \left | \left| \boldsymbol{x}^s - \boldsymbol{x}_r \right | \right |_2 = \sqrt{(x^s - x_r)^2 + (y^s - y_r)^2 + (z^s - z_r)^2}$ represents the euclidean distance between receiver $r$ and satellite $s$ consisting of the respective Cartesian coordinates. In addition, $\delta t_r$ and $\delta t^s$ represent the receiver and the satellite clock error respectively, $\delta_{\text{ion}}^s$ and $\delta_{\text{tro}}^s$ the atmospheric error influences and finally $\varepsilon_r^s$ as the unmodeled, uncorrelated error terms for each pseudorange measurement. Following up on this, BSSD performs a differencing between respective observations. For the generic satellites $s_1$ and $s_2$ this yields:

\begin{equation}
    \label{equ:sd}
    \Delta \rho_{r}^{s_1, s_2} = \rho_r^{s_1} - \rho_r^{s_2}.
\end{equation}

Given \cref{equ:pseudorange_model} we obtain:

\begin{align}
    \label{equ:sd_long}
    \Delta \rho_{r}^{s_1, s_2} &=  d_r^{s_1} - d_r^{s_2} + c(\delta t_r - \delta t_r) - c(\delta t^{s_1} - \delta t^{s_2}) + \delta_{\text{ion}}^{s_1} - \delta_\text{ion}^{s_2} + \delta_{\text{tro}}^{s_1} - \delta_\text{tro}^{s_2} + \boldsymbol{\sigma}_r^{s_1, s_2} \\
    &= d_r^{s_1} - d_r^{s_2} + c\Delta \delta t^{s_1, s_2} + \Delta\delta_{\text{ion}}^{s_1, s_2} + \Delta\delta_{\text{trop}}^{s_1, s_2} + \boldsymbol{\sigma}_r^{s_1, s_2} \;.
\end{align}

The obtained model corresponds to the hyperbolic geometrical model introduced in \cref{ssec:hybrid_grid_filter}. Besides the elimination of the receiver-specific clock error, the observable combination mainly affects the unmodeled error terms. Due to the observation combination of BSSD and according to the error propagation with the assumption of $\sigma_r^s = \sigma_r^{s_1} = \sigma_r^{s_2}$ the influence of the stochastic error terms is given as follows \citep{Odijk_gnss_handbook_differential_positioning_2017}:

\begin{equation}
    \sigma_r^{s_1, s_2} = \sigma_{\text{SD}} = \sqrt{2} \sigma_r^{s_1, s_2}.
\end{equation}

Concerning the parametrization of $\sigma_r^s$ a variety of procedures are applicable. At first, generic values according to system specifications can be set, e.g. using the GPS Signal in Space pseudorange error budget of the Standard Positioning Service \cite{Schwarzbach2020SingleDiff}. Furthermore, an empirical parameter setting based on available measurements can be obtained \cite{Groves2020LikelihoodRanging}. For simplification, we stick to the former possibility and assume a mean-free Gaussian distribution given a line-of-sight (LOS) standard deviation of $\sigma_r^s = \SI{7.8}{\meter}$, leading to $\sigma_{SD} = \sqrt{2}\sigma_r^s$.

In recent literature, a variety of algorithmic 3DMA approaches with a holistic focus on adequate NLOS handling based on 3D map data are discussed, e.g. \citep{Groves2020LikelihoodRanging}. As this not the sole focus of this work, we want to present a more general approach, which can also be implemented without the knowledge of building information or other spatial data apart from the utilized discrete state space.

BSSD approaches are implemented by selecting a pivot satellite assumed to be LOS, which is then differenced from all other observations. Depending on the estimated LOS/NLOS state of the observation, different stochastic models are applied. As presented in \citep{Schwarzbach2020SingleDiff}, a full set differencing between all observed satellites is also applicable, resulting in a symmetric error distribution. An example given the static dataset presented in \cref{sec:data} is given in \cref{fig:pseudorange_residuals}. 

This empirical survey reveals that the differencing across all satellites produces characteristic effects:

\begin{itemize}
    \item Case 1: The differencing of LOS measurements leads to a mean-free residual distribution.
    \item Case 2 and 3: Mixture densities for combined LOS/NLOS differencing depending on the associative law and if the LOS satellite is differenced from the NLOS one or vice versa. Case 2 corresponds to a positive mean, case 3 to a negative.
    \item Case 4: Measurement outliers, which do not fit one of the former.
\end{itemize}

In order to account for these effects, the modeling of a Gaussian mixture model (GMM) is suggested, allowing the exploitation of the full set differencing. In our case, a $3$ component GMM is applied, whose parametrization is presented in \cref{sec:data}. At first, a visibility prediction of observed satellites based on the building boundary information \citep{Groves2020LikelihoodRanging} is performed. Subsequently, full set differencing for all satellites $i$ and $j$ ($i \neq j$) is performed. The selection of the GMM component for sampling corresponds to the cases listed above:

\begin{itemize}
    \item $i$ and $j$ are predicted as LOS satellites, $\text{GMM}_1$ is selected;
    \item $j$ is predicted LOS and $i$ is predicted NLOS, $\text{GMM}_2$ is selected;
    \item $i$ is predicted LOS and $j$ is predicted NLOS, $\text{GMM}_3$ is selected;
    \item For any other case, the differenced observations are ignored.
\end{itemize}

\subsubsection{Terrestrial Measurement model}\label{ssec:terrestrial_meas}

In accordance with \cref{ssec:terrestrial_systems}, a variety of technologies and accompanying measurement principles for positioning is available. The empirical validation of the method presented in \cref{sec:data} introduces an UWB real-time location system based on two-way ranging, which is a realization of RTT. Respective implementations are detailed in \citep{LianSang2019TwoWayRanging}, specified in the UWB standard and therefore available in commercial products. The underlying geometrical model corresponds to the distance measurement described in \cref{ssec:measurement}. 

Unlike 3DMA approaches and due to the volatile nature of the propagation between terrestrial stationary reference points and mobile devices, geometry based NLOS identification cannot be easily applied. For this task however, a variety of algorithmic mitigation strategies exist \citep{Guvenc2007UWBNLOSSurvey}.  The probabilistic error mitigation favored in this work uses a Likelihood mixture model, as originally presented in \citep{Fox2001ParticleFilterMixture}. The mixture Likelihood uses a set of proposal distributions to sample from and combines them using a mixture ratio $\phi$. For the example, of two proposal distributions $\mathcal{D}_1$ and $\mathcal{D}_2$ the mixture Likelihood $\mathcal{D}_{\text{ML}}$ yields:

\begin{equation}
    \mathcal{D}_{\text{ML}} = (1 - \phi) \mathcal{D}_2 + \phi \mathcal{D}_1
\end{equation}

The parametrization is based on a statistical analysis of UWB ranging error distributions presented in \citep{Schwarzbach_statistical_evaluation_2021}. Essentially, UWB observations and accompanying error influences $\epsilon$ are categorized into three propagation scenarios: 

\begin{enumerate}
    \item LOS reception following a mean-free Gaussian distribution, 
    \item NLOS reception leading to a right-skewed residual distribution or
    \item Measurement outliers and failures.
\end{enumerate}

The individual error magnitudes and corresponding occurrence probabilities are further discussed in \cref{sec:data}.



\subsubsection{Motion model}
The motion model is calculated as described in \cref{ssec:prediction}. Basically, there are two ways of implementing the motion model. On the one hand, model-based assumptions about the motion of the object to be located can be formulated. This usually includes a model assumption of the velocity, associated with a corresponding uncertainty. On the other hand, the use of dynamic data of the object is possible. A practical implementation is the odometry motion model \citep{thrun2005probabilistic}, which uses the velocity of the object (speed over ground) and its heading. 

\begin{figure}[htb!]
    \centering
    \includegraphics[width=.95\linewidth]{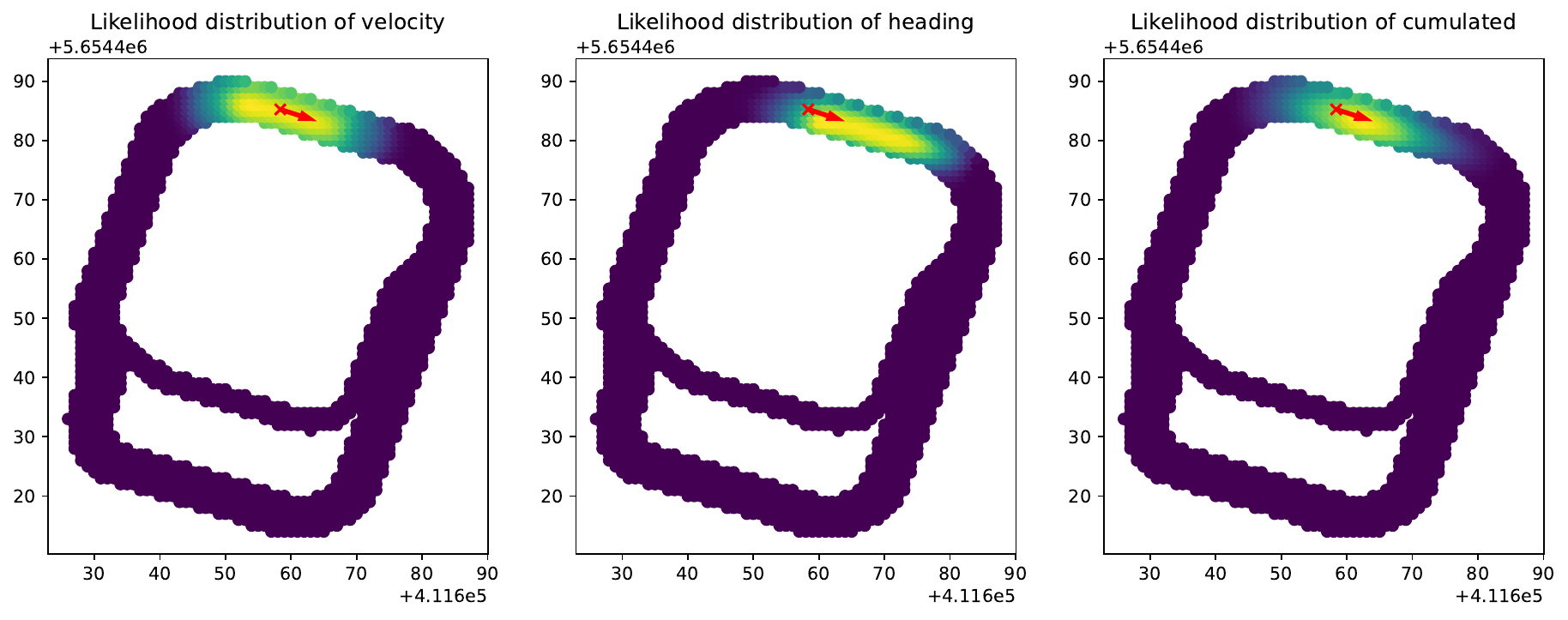}
    \caption{Motion likelihood distribution for the most probable current candidate point (red dot). Motion vector (red) is added for clarity.}
    \label{fig:likelihood_motion}
\end{figure}

The information necessary for applying the odometry motion model can be obtained from a variety of sensory inputs. In this work, we simply use the velocity and heading information estimated from the reference receiver, which is introduced in \cref{sec:data}. In unison with the measurement step, the (expected or assumed) quality of information for the prediction step can be modeled in a probabilistic manner (cf.~\cref{equ:grid_predict_distance} or \ref{equ:grid_predict_speed_heading}). The results of the prediction sampling step for the velocity, heading and combined information is depicted in \cref{fig:likelihood_motion}. 

\section{Data Acquisition and Evaluation}
\label{sec:data}

\subsection{Methodology}
In order to evaluate the performance of the implemented approach, test runs are performed on the testbed for automated and connected driving in Dresden, Germany.  Those runs consist of static and dynamic scenarios. The static test runs are conducted at fixed locations, which were geo-referenced with a survey-grade GNSS setup. The data from the static samples are used to analyze the distribution of GNSS pseudorange residuals. The parameters of the resulting multi-modal distribution are then applied in the parametrization of the final hybrid Grid Filter. This filter is evaluated in a dynamic driving test. In \cref{tab:test_cases} the conducted scenario, used sensors and reference receivers as well as the outcomes of each scenario are summarized.

\begin{figure}[htb!]
    \centering
    \begin{subfigure}[b]{0.2821\textwidth}
    \centering
    \includegraphics[width=1\linewidth]{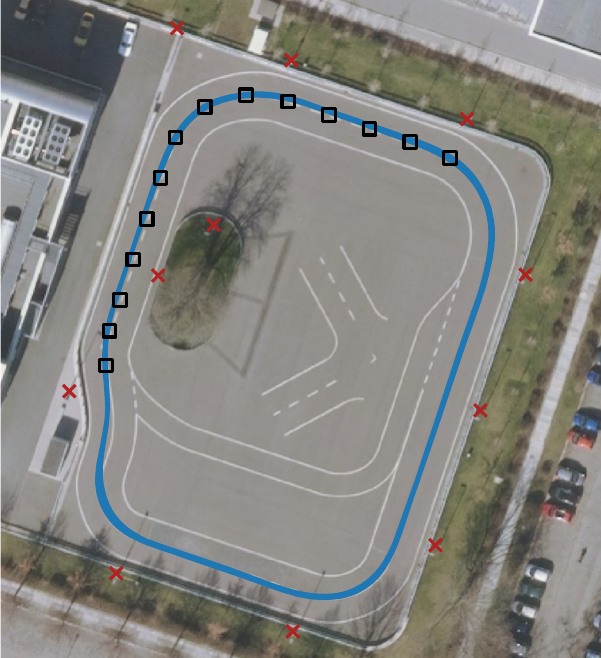}
    \caption{}
    \label{fig:reference_positions}
    \end{subfigure}
    \centering
    \hspace{1cm}
    \begin{subfigure}[b]{0.55\textwidth}
    \centering
    \includegraphics[width=1\linewidth]{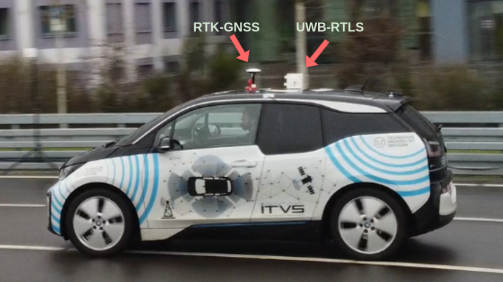}
    \caption{}
    \label{fig:i3}
    \end{subfigure}
    \caption{\textbf{(a)} Visualization of terrestrial pseudolites (red) and surveyed reference points: static points (black) and dynamic trajectory (blue); \textbf{(b)} Test vehicle with sensor setup. The antenna offset was corrected during post-processing using the known baseline between both systems.}
    \label{fig:setup}
\end{figure}

The applied accuracy metric $\mathcal{Q}$ corresponds to the three-dimensional L2-norm $\lVert \cdot \rVert_2$ between the reference $\boldsymbol{\mathrm{x}}_{\text{ref}, k}$ and estimated $\hat{\boldsymbol{\mathrm{x}}}_k$ positions at each time step $k$:

\begin{equation}
    \label{eq:rmse}
    \mathcal{Q} = \lvert\lvert \boldsymbol{\mathrm{x}}_{\text{ref}, k} - \hat{\boldsymbol{\mathrm{x}}}_k \rvert\lvert_2
\end{equation}

\begin{table}[htbp!]
\centering
\caption{Test cases on testbed for automated and connected driving.}
\begin{tabular}{@{}llllll@{}}
\toprule
            & Test case                 & Description                                            & Sensors         & Outcomes & Reference                               \\ \midrule
\multirow{2}{*}{I}  & \multirow{2}{*}{static}  & \multirow{2}{*}{14 points on testbed} & u-blox F9P GNSS receiver      & GNSS PR Residuals     &   \multirow{2}{*}{Leica GS15}                \\
                    &                          &                                                        & ZigPos UWB RTLS & UWB Ranging Residuals        &        \\
                    &                          &                                                        &          &                             &          \\
\multirow{2}{*}{II} & \multirow{2}{*}{dynamic} & \multirow{2}{*}{Evaluation run}                     &  u-blox F9P GNSS receiver      & \multirow{2}{*}{Algorithmic evaluation  } & \multirow{2}{*}{NovaTel PwrPak7} \\
                    &                          &                                                        & ZigPos UWB RTLS &                                        &    \\
 \bottomrule                                 
\end{tabular}
\label{tab:test_cases}
\end{table}

\subsection{Data Set}
  For GNSS data collection, an u-blox F9P receiver was used on the vehicle. The recorded data include pseudorange estimates of GPS L1, Galileo E1 and Glonass L1 FDMA.  As terrestrial radio system, a UWB system by Zigpos GmbH was used, which features two-way ranging based on DecaWave chip sets. For this purpose, 11 UWB anchors were placed and geo-referenced in order to be integrated with GNSS.  To enable a comprehensive performance and error analysis, all measurements were referenced. The offset between UWB and GNSS antenna is corrected using the known baseline parameters between both.  The static scenario consists of 4038 measurement epochs on 14 points on the testbed (cf,~\cref{fig:setup}). The aim is to analyze error distributions of the applied sensor systems given operation environment. For the static scenarios, a survey-grade Leica GS15 with VRS RTK correction data was used as reference receiver. 

The dynamic scenario consists of a test run using a test vehicle depicted in \cref{fig:setup}. The data set consists of 2066 raw measurement epochs, of which 1411 include GNSS and 655 include UWB data. The UWB measurement rate was set comparatively low to showcase the influence of terrestrial augmentation for GNSS. The reference was recorded using a Novatel PwrPak 7 with VRS RTK corrections.

\subsection{Evaluation}

\begin{figure}[htb!]
    \centering
    \begin{subfigure}[b]{0.32\textwidth}
    \centering
    \includegraphics[width=1\linewidth]{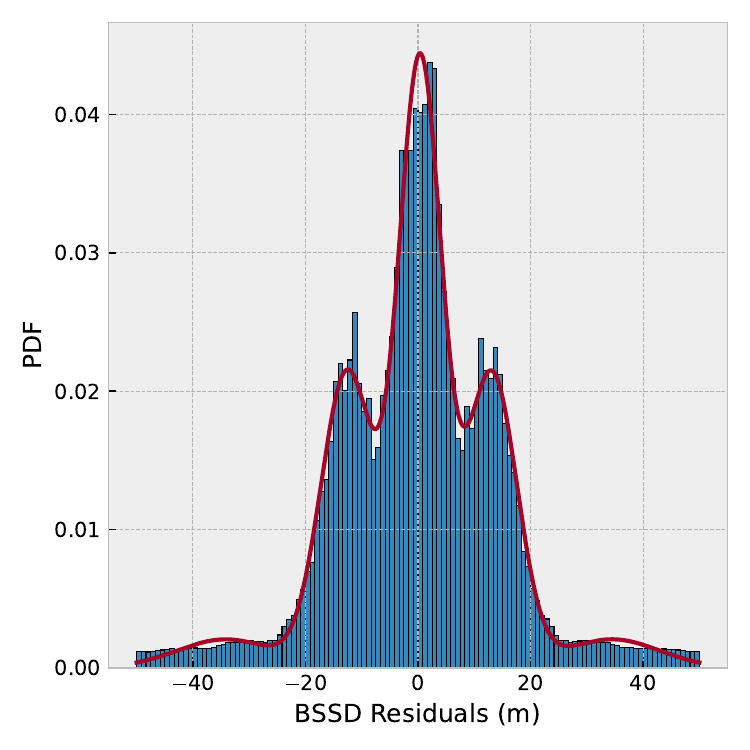}
    \caption{}
    \label{fig:pseudorange_residuals}
    \end{subfigure}
    \centering
    \begin{subfigure}[b]{0.32\textwidth}
    \centering
    \includegraphics[width=1\linewidth]{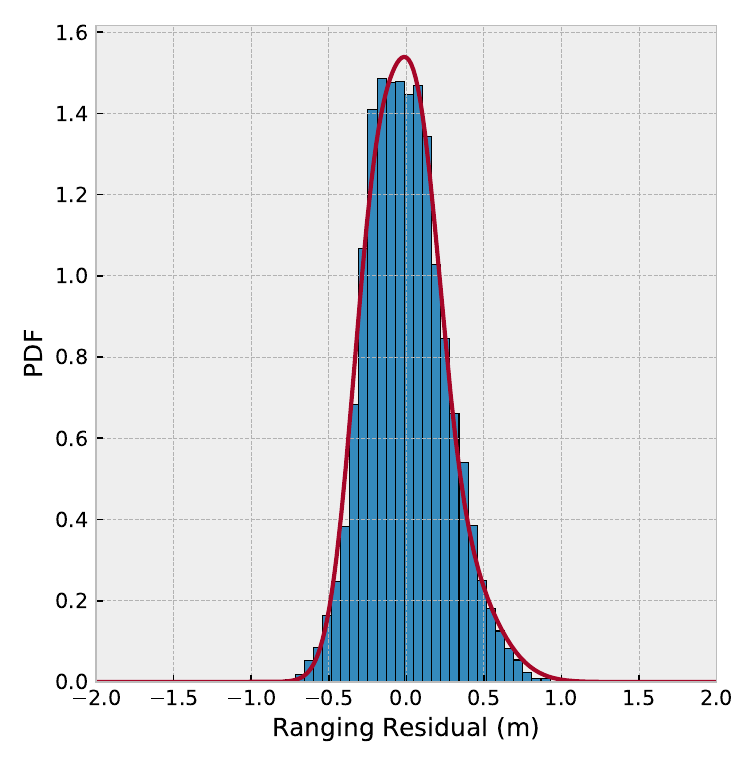}
    \caption{}
    \label{fig:ranging_residuals}
    \end{subfigure}
    \begin{subfigure}[b]{0.275\textwidth}
    \centering
    \includegraphics[width=1\linewidth]{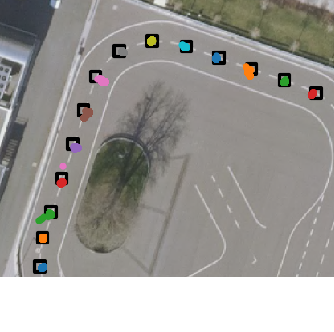}
    \caption{}
    \label{fig:qual_result_static}
    \end{subfigure}
    \centering
    \caption{Histogram (blue) of the error distributions of the underlying sensor information: \textbf{(a)} Clock corrected GNSS pseudorange residuals and \textbf{(b)} UWB ranging residuals (kernel density estimation is given in red). \textbf{(c)} Orthofoto of qualitative positioning results for the static scenario.}
    \label{fig:residuals}
\end{figure}

At first, the static scenario and more specifically the quality of the available sensor information are evaluated. For this purpose, \cref{fig:residuals} includes a histogram of the full set differencing approach in \cref{fig:pseudorange_residuals} and the UWB ranging residuals in \cref{fig:ranging_residuals}.

As aforementioned, the full set satellite differencing yields a symmetric residual distribution. As described in \cref{ssec:gnss_meas}, the resulting mixture distribution can be approximated as GMM using a variable amount of $C$ Gaussian components associated with different weights $\omega_c$, means $\mu_c$ and variances $\Sigma_c$:

\begin{equation}
    P \sim \sum_c^C  w_c \cdot \mathcal{N}(\mu_c, \Sigma_c)
    \label{eq:gmm}
\end{equation}

For the surveyed data, the underlying GMM is estimated using the Python library scikit-learn \citep{scikit-learn} assuming $C = 4$ components. The respective values for the BSSD GMM components are summarized in \cref{tab:gmm}. The estimated standard deviations of each sensor system is then applied in the parametrization of the measurement update step of the respective sensor. 

\begin{table}[htbp!]
\renewcommand{\arraystretch}{1.2}
\centering
\caption{Estimated GMM parameter for the BSSD residuals obtained from the static scenario.}
\begin{tabular}{@{}cccc@{}}
\toprule
  & $\omega_c$ & $\mu_c$ & $\Sigma_c$ \\ \midrule
$1$ & $0.42$           &    $0.25$    &     $13.06$       \\
$2$ & $0.24$           &    $13.09$     &    $20.37$        \\
$3$ & $0.24$           &    $-12.61$    &     $21.05$       \\
$4$ & $0.01$           &    $-0.3$     &    $142.89$        \\ \bottomrule
\end{tabular}
\label{tab:gmm}
\end{table}

In addition, the accuracy of the UWB ranging measurements is evaluated as shown in \cref{fig:ranging_residuals}. Unlike the theoretical assumptions formulated in \cref{ssec:terrestrial_meas}, the data does not contain any NLOS measurements, therefore this part of the mixture can be neglected. This effect can be reasoned due to the advantageous propagation conditions on the testbed. Apart from vegetation, LOS propagation is present. The parameters of the normal distribution shown in \cref{fig:ranging_residuals} are:

\begin{equation}
    \varepsilon_{\text{UWB}} \sim \mathcal{N}(\mu_{{UWB}}, \sigma_{\text{UWB}}^2) \quad \quad \text{with} \quad \quad \mu_{\text{UWB}} = \SI{0.05}{\meter} \quad \text{and} \quad \sigma_{\text{UWB}} = \SI{0.31}{\meter} \; . \vspace{0.5cm}
\end{equation}

In addition, a variety of measurement outliers occurred, accumulating to a total of approximately $\SI{10}{\percent}$. Therefore, mixture ratio for the mixture Likelihood distribution can empirically be set to $\phi = 0.9$. 

Applying the presented hybrid Grid Filter with these parameter settings for GNSS and UWB observations to the surveyed data yields the qualitative positioning results shown in \cref{fig:qual_result_static}. As summarized in \cref{tab:statistics} a mean $\bar{\mathcal{Q}} = \SI{0.64}{\meter}$ respectively a median $\tilde{\mathcal{Q}} = \SI{0.62}{\meter}$ positioning error is achieved.
 
A summary of descriptive statistics for the error distributions for the static and dynamic scenario is given in \cref{tab:statistics}. This includes the average error $\bar{\mathcal{Q}}$ (root mean square error), the median error $\tilde{\mathcal{Q}}$ and the error variance $\sigma_{\mathcal{Q}}^2$. In addition, the $\sigma$ quantiles and the $25, 50, 75$ percentiles of the error distributions are given.

\begin{figure}[htbp!]
    \centering
    \begin{subfigure}[b]{0.45\textwidth}
    \centering
    \includegraphics[width=\linewidth]{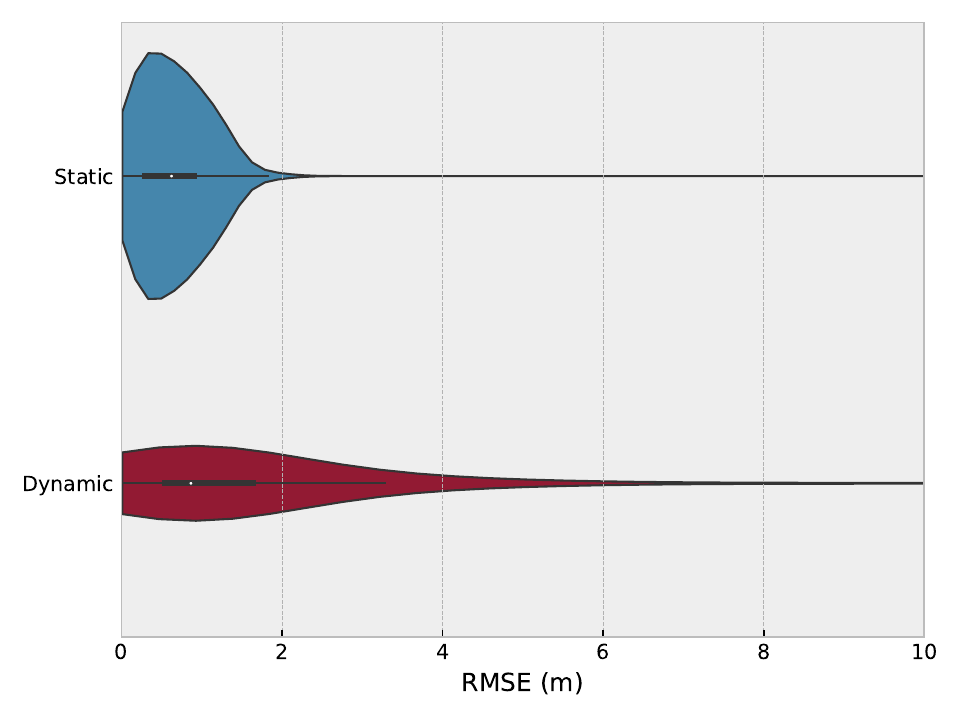}
    \caption{}
    \label{fig:violin}
    \end{subfigure}
    \centering
    \begin{subfigure}[b]{0.45\textwidth}
    \centering
    \includegraphics[width=\linewidth]{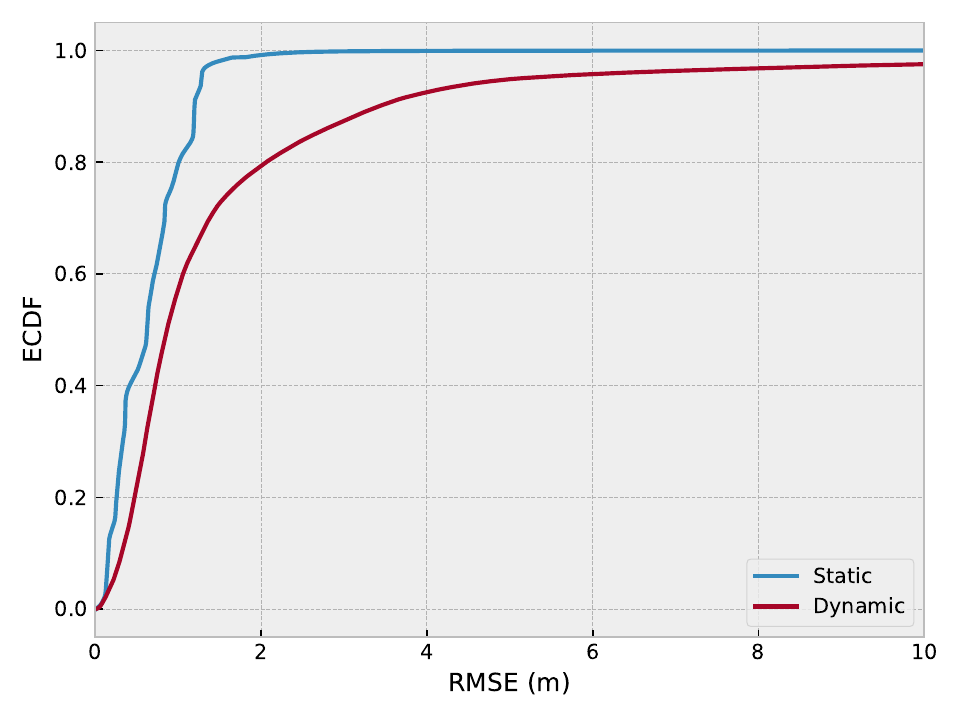}
    \caption{}
    \label{fig:ecdf}
    \end{subfigure}
    \caption{Statistical evaluation of the root squared error for both the static (blue) and dynamic (red) scenario: \textbf{a)} Violinplot and  \textbf{b)} empirical cumulative density function (ECDF).}
    \label{fig:statistic}
\end{figure}

Furthermore, the proposed method and parameter settings are also applied to the dynamic data set. A quantitative error analysis for both scenarios is given \cref{fig:statistic}. In general, the hybrid Grid Filter applied to the dynamic scenario achieves less positioning accuracy, accumulating a mean $\bar{\mathcal{Q}} = \SI{1.62}{\meter}$ respectively a median $\tilde{\mathcal{Q}} = \SI{0.84}{\meter}$ positioning error associated with a higher error variance.  This can be accounted to the fact, that the static case allows for a convergence on the correct position over a longer period of time due to the static motion model. The sequential motion update does not induce as much noise due to the constant position behavior of the system. This is further reflected in the comparably low variance of the positioning error with $\sigma_{\mathcal{Q}}^2 = \SI{0.21}{\meter}^2$. Apart from a single outlier, observable in \cref{fig:qual_result_static} (pink), which rapidly converges, the position results are comparably stable.

In contrast, the dynamic motion model allows for a smoothing of the trajectory, but does not always prevent jumps in the state space due to observation outliers. One approach to compensate for this would be a more restrictive parametrization of the motion model. However, a more loosely configured motion model facilitates recovery of biased measurements over multiple epochs. This dichotomy can be resolved by adaptive parametrization based on statistic evaluation of measurement residuals, as provided by integrity monitoring techniques such as in \citep{Zhong20223DMAOutlierDetection}. 


\begin{table}[htbp!]
\renewcommand{\arraystretch}{1.4}
\centering
\caption{Quantitative evaluation of the static and dynamic scenario.}
\begin{tabular}{@{}cccccccccc@{}}
\toprule
\multirow{2}{*}{Measure} & $\bar{\mathcal{Q}}$    & $\tilde{\mathcal{Q}}$  & $\sigma_{\mathcal{Q}}^2$ & \multicolumn{3}{c}{Quantile $[ \SI{}{\meter} ]$          }                         & \multicolumn{3}{c}{Percentile $[ \SI{}{\meter} ]$} \\ 
                         & $[ \SI{}{\meter} ]$ &  $[ \SI{}{\meter} ]$ & $[ \SI{}{\meter} ]^2$               & $\sigma$ & $2\sigma$ & $3\sigma$ & $25$          & $50$          & $75$         \\ \midrule
Static                   &   $0.64$      &   $0.62$      &    $0.21$                                      &   $0.82$                    &     $1.28$                   &      $1.72$                  &   $0.28$          &     $0.62$        &    $0.91$       \\
Dynamic                  &   $1.62$      &  $0.84$       &       $6.56$                                   &      $1.31$                 &     $4.82$                   &         $15.60$               &  $0.54$           &    $0.86$         &     $1.64$       \\ \bottomrule
\end{tabular}
\label{tab:statistics}
\end{table}

\section{Conclusion}
In conclusion, we identify great potential in the unified integration of both GNSS observations and radio-based terrestrial relations using a hybrid Grid Filter. This is exemplified by a test run on an automotive test bed, in which GNSS, UWB and motion data are fused using an implementation of such a filter. The parametrization of the measurement model is done by applying the measurement residuals extracted from a static test, which was conducted within the same reception environment. 
Furthermore, 3D models are applied to account for NLOS reception. 
Our test shows the general feasibility of the approach, resulting in a RMSE of \SI{1.62}{m} in the surveyed dynamic case. The results however also indicate a lot of potential for future work, including:
\begin{itemize}
    \item Inclusion of additional sensor systems 
    \item Application of additional geometric relations, e.g. terrestrial AoA or TDoA
    \item Fine-tuning of measurement models, integrated adaption of advanced 3DMA approaches as well as outlier detection for both GNSS and terrestrial observations
\end{itemize}

In addition to these algorithmic potentials, we also recognize different application potentials to further evaluate the presented method, e.g. by focusing on indoor-outdoor localization scenarios or possible technology hand-over scenarios. Additionally, a data fusion with opportunistic signals within intelligent transportation systems can be addressed.

\section*{ACKNOWLEDGEMENTS}

\begin{tabular}{p{12.cm} p{3cm} p{3cm}}
The authors want to thank ZigPos GmbH for supporting the UWB data acquisition.
This work has been funded by the German  Federal Ministry for Digital and Transport (BMDV) following a resolution of the German Federal Parliament within the projects IDEA (FKZ: 19OI22020C). & 
\hspace{1 cm} \raisebox{-0.8\height}{\includegraphics[width=0.6\linewidth]{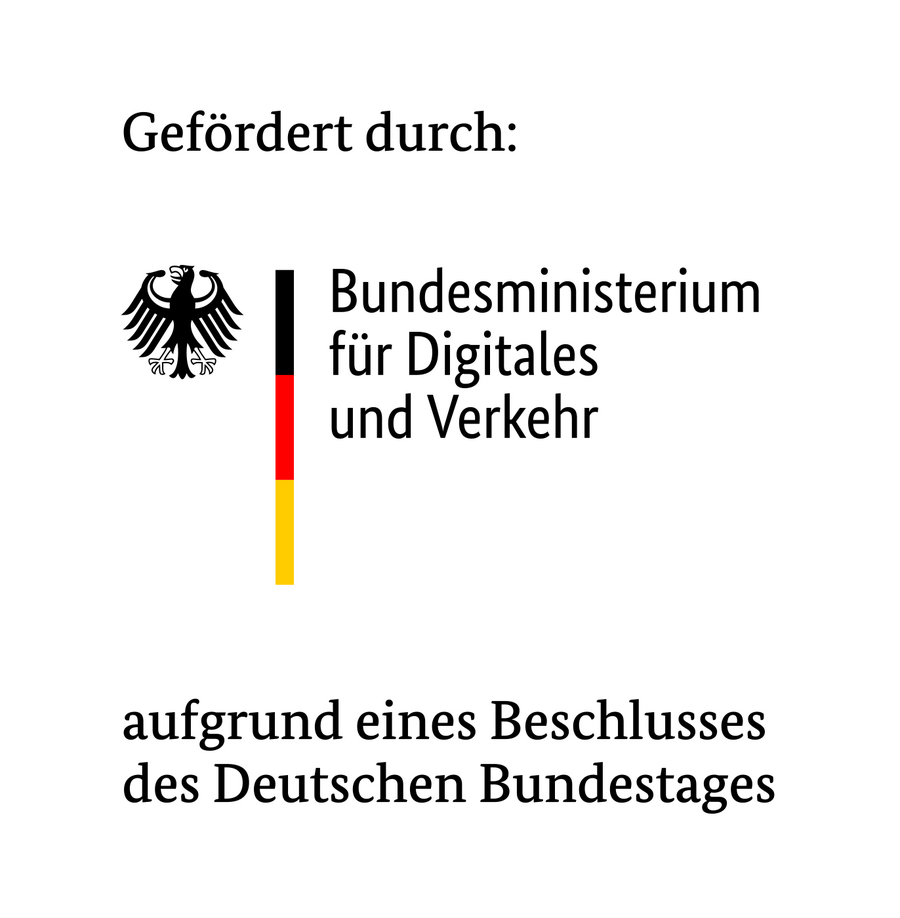}} & \hspace{-0.5 cm} \raisebox{-0.8\height}{\includegraphics[width=0.7\linewidth]{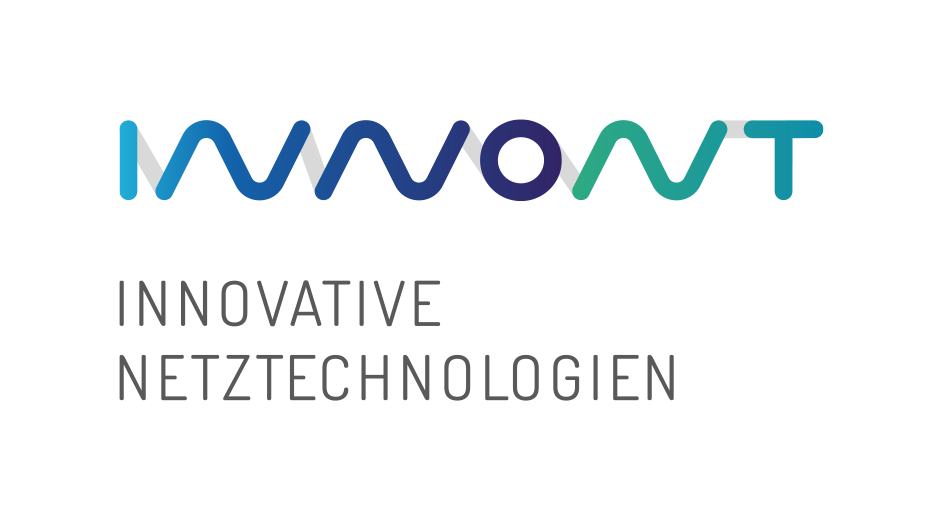}}
\end{tabular}

\bibliographystyle{apalike}
\bibliography{lit.bib}

\end{document}